\documentclass[journal,showpacs]{IEEEtran}

\usepackage{amsmath,amssymb,graphicx}
\usepackage{multirow}
\usepackage{epsf}
\usepackage{subfigure}
\usepackage{hyperref}

\usepackage{CJK}
\usepackage{amsmath}
\usepackage{indentfirst}
\usepackage{graphicx}
\usepackage{float}
\usepackage{comment}
\newtheorem{theorem}{Theorem}

\newtheorem{definition}[theorem]{Definition}

\begin{document}

\title{Secret key distillation across a quantum wiretap \\channel under restricted eavesdropping}

\author{Ziwen Pan$^{*}$, Kaushik P. Seshadreesan$^{\dagger}$, William Clark$^{\#}$, Mark R. Adcock$^{\#}$, \\
Ivan B. Djordjevic$^{*}$, Jeffrey H. Shapiro$^{\ddag}$, Saikat Guha$^{\dagger}$
\thanks{$^{*}$Department of Electrical \& Computer Engineering,
        the University of Arizona, 1230 E Speedway Blvd, Tucson, AZ 85719}
        
        \thanks{$^{\dagger}$College of Optical Sciences,
        the University of Arizona, 1630 E University Blvd, Tucson, AZ 85719}
\thanks{$^{\#}$General Dynamics Mission Systems, 8220 East Roosevelt Street, Scottsdale, AZ 85257, USA}

\thanks{$^{\ddag}$Research Laboratory of Electronics, Massachusetts Institute of Technology, Cambridge, Massachusetts 02139, USA}}

\maketitle

\begin{abstract}
The theory of quantum cryptography aims to guarantee
unconditional information-theoretic security against an omnipotent eavesdropper. In many practical scenarios, however, the assumption of
an all-powerful adversary is excessive and can be ded considerably. In this paper we study
secret key distillation across a lossy and noisy quantum wiretap channel between Alice and
Bob, with a separately parameterized realistically lossy quantum channel to
the eavesdropper Eve. We show that under such restricted eavesdropping, the key rates
achievable can exceed the secret key distillation capacity against an unrestricted 
eavesdropper in the quantum wiretap channel. Further, we show upper bounds on the key rates based on the relative entropy of entanglement. This simple restricted eavesdropping
model is widely applicable, e.g., to free-space quantum optical communication, where realistic
collection of light by Eve is limited by the finite size of her optical aperture. Future work will
include calculating bounds on the amount of light Eve can collect under various realistic
scenarios.
\end{abstract}



\section{Introduction}

\IEEEPARstart{Q}{uantum} key distribution (QKD) theoretically promises unconditional security in the physical layer. Bennett and Brassard~\cite{bennet1984proceedings} developed the first QKD protocol (BB84), whose security is guaranteed by the no-cloning theorem of quantum mechanics and one-time-pad encryption~\cite{SPCDLP09}. Numerous commercial products are available today that implement variants of the decoy-state BB84 (DS-BB84) protocol~\cite{LMC05} based on polarization qubits encoded in weak coherent state pulses.

While discrete variable (DV) protocols such as the DS-BB84 showcase the power of quantum cryptography, the key rates achievable are low and the systems are difficult to integrate with existing telecommunication systems. Hence, nowadays there is a thrust to develop QKD systems that can overcome these challenges. In this regard, continuous variable (CV)-QKD schemes, e.g., based on coherent laser light and heterodyne detection are being viewed as viable solutions~\cite{LPFPP18, DL15}.

Traditionally, the security proofs of quantum key distribution against a wiretapping adversary, an eavesdropper Eve, assume that Eve can perform any operation allowed by the laws of quantum physics on the transmitted light, and has access to all the light that is lost in transmission. However, this is not the case in some realistic applications, especially in free-space communication channel where it is reasonable to consider a potential Eve who is  restricted in her information collection capabilities. 

\begin{figure}[H]
\centering
{\includegraphics[height=2.3pc]{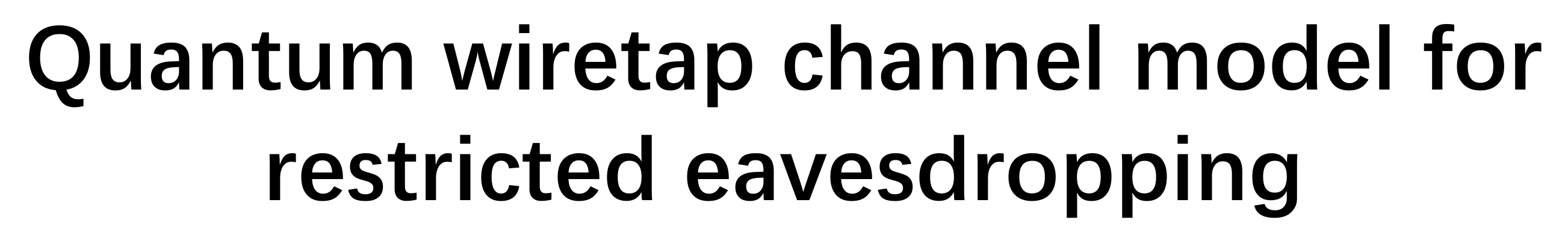}\\
\includegraphics[width=0.361\textwidth]{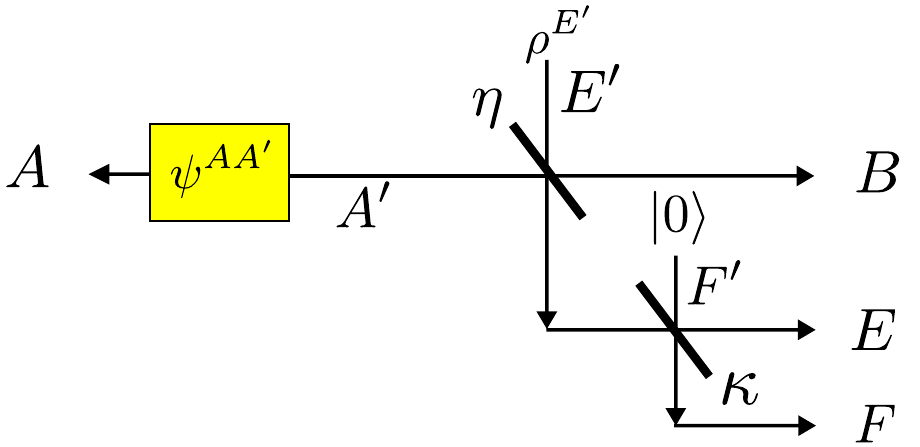}}\\
\caption{Entanglement-based model for quantum communication over a wiretap channel of transmissivity $\eta$ from Alice to Bob under restricted eavesdropping. Alice prepares an entangled pure state $|\psi\rangle^{AA'}$ and sends $A'$ through the channel to Bob while retaining $A$ with herself. Bob receives $B$ at the output of the channel. The restriction on the eavesdropper Eve is modeled by a pure loss beam splitter of transmissivity $\kappa$. Eve is shown to inject a state $\rho^{E'}$ into the channel to Bob, which could be a vacuum state (passive attack) or a thermal state (active attack). Then Eve collects $E$ and can perform any operations allowed by the physics law  to eavesdrop. $F$ is considered lost in the process of Eve's collecting action and thus inaccessible to any party.	\label{fig1}}
\end{figure}

In this work, we present a secure key rate analysis for a secret key distillation scheme over a quantum wiretap channel from a sender Alice to receiver Bob, where the eavesdropper Eve is restricted to receive only a fraction of the photons lost in transmission as shown in Fig.~\ref{fig1}. As a result, some of the light is rendered inaccessible to any of the parties involved and lost to the environment. Such a restriction is widely applicable, e.g., in optical wireless communication~\cite{UH14}, where a realistic Eve would be limited by the size of the aperture of her receiver, or be forbidden to collect light from an exclusion zone around Alice-to-Bob line of sight. We consider both passive eavesdropping where Eve injects the vacuum state into the channel, and an instance of active eavesdropping, where Eve injects a thermal state into the channel. \footnote{Note that other possible restrictions on Eve include, e.g., a noisy quantum memory~\cite{WST08} of finite size and coherence time, and a noisy communication channel where some of the noise is well characterized and provably of non-adversarial origin, which does not benefit Eve either. However, in this work, we focus on the restricted light collection capability of Eve.} The analysis would include both direct and reverse reconciliation~\cite{grosshans2002reverse,grosshans2003virtual} as they improve differently under restricted eavesdropping.

The main findings of our analysis include the following: 
\begin{itemize}
	\item{Invoking the Hashing inequality~\cite{devetak2005distillation} for secret key distillation over a communication channel with one-way public discussion, we write down lower bounds on asymptotic achievable key rates under the restricted eavesdropping model. We show that these key rates can exceed the direct transmission capacity of the channel in a traditional unrestricted eavesdropping model. This implies both higher key rates as well as longer transmission distances.}
	
	\item{We provide upper bounds on the key rates under the restriction on Eve based on the relative entropy of entanglement ($E_\textrm{R}$). In the case of the pure loss channel, the $E_\textrm{R}$ upper bound closely matches the achievable rate with heterodyne detection and reverse reconciliation.}
	
	\item{We present a comparison of a CVQKD protocol based on Gaussian modulated coherent states and heterodyne detection, and the DS-BB84 DVQKD protocol based on polarization encoding of weak coherent states, under both passive and active  restricted eavesdropping models.}
\end{itemize}
	
	The paper is organized as follows: Sections~\ref{LBs} and~\ref{UBs} describe the methods employed in establishing the achievable key rates and the upper bounds on secret key distillation under restricted eavesdropping. These methods apply to passive as well as active eavesdropping, and direct and reverse information reconciliation schemes. In Section~\ref{results}, we present the results obtained by applying these methods to the entanglement-based model for secret key distillation based on the two-mode squeezed vacuum state and heterodyne detection. In Section~\ref{comp}, we present results comparing achievable key rates (under the restricted eavesdropping model) in CVQKD with Gaussian modulated coherent states and heterodyne detection, and in DS-BB84 protocol. We conclude with a summary in Section~\ref{concl}.

\section{Achievable rates for secret key distillation under restricted eavesdropping}\label{LBs}

Consider the entanglement-based model for bipartite secret key distillation shown in Fig.~\ref{fig1}, where Alice prepares an asymptotically large number of independent and identically distributed (i.i.d.) copies of an entangled pure state $\psi^{AA'}$ and transmits the $A'$ systems through a wiretap channel. In this limit, the optimal attack for an eavesdropper Eve is known to be a collective attack~\cite{RC09}, namely wherein she performs an identical symbol-by-symbol active attack on each transmission, resulting in i.i.d. copies of a bipartite state $\rho^{AB}$ being shared between Alice and Bob.

Let us first recall key rate analysis for secret key distillation against an unrestricted Eve ($\kappa=1$ in Fig.~\ref{fig1}).

\subsection{Unrestricted Eve}\label{LBunrestrictedEve}

In traditional security analysis for key distillation over a quantum wiretap channel, where the eavesdropper Eve is assumed to have access to all the light that is lost in transmission, she holds the full purification of the state $\rho^{AB}$. In other words, Eve holds a quantum system $E$ such that the systems $A$, $B$ and $E$ are in a pure state $|\psi\rangle^{ABE}$ satisfying $\rho^{AB}=\operatorname{Tr}_E(\psi^{ABE})$. In the case of a thermal noise channel with loss, Eve holds the purification of both the state $\rho^{E'}$ she injected into the channel as well as the output of the channel to Bob, namely a pure state $|\psi\rangle^{ABER}$, $R$ being the purifying system of Eve's input $\rho^{E'}$. Note that the systems $ER$ in this case can be thought of as one joint purifying system $E$. Hence the system $E$ in subsequent discussion in this section also includes the case of the thermal noise channel.

When either Alice or Bob performs a measurement, Devetak and Winter~\cite{devetak2005distillation} proved an achievable rate ($K_{\textrm{CQQ}}$) for secret key distillation from the resulting state with one-way public classical communication assistance---a result known as the Hashing inequality. Here CQQ stands for ``Classical-Quantum-Quantum", indicating that either Alice or Bob has performed a measurement on her/his quantum system, while the other and Eve are yet to measure their respective quantum systems. For example, in a reverse reconciliation scheme, Bob performs a measurement, while Eve and Alice retain their quantum systems unmeasured. (Likewise, CCQ stands for ``Classical-Classical-Quantum", which indicates that both the communicating parties have performed measurements. $K_\textrm{CCQ}$ reflects the corresponding achievable key rate, and will be discussed in Section~\ref{comp} in the context of QKD protocols.)

The Hashing lower bound on the key rate for direct reconciliation, namely when Alice measures system $A$ to give rise to a classical outcome $X$ that is publicly communicated to Bob is given by
 \begin{align}
 K_\rightarrow(\rho)\geq I(X;B)_\omega-I(X;E)_\omega.
 \end{align}
Here the state $\omega^{XBE}$ is
\begin{equation}
\omega^{XBE}=\int_x\!dx\,p(x)|x\rangle\langle x|^X\otimes\rho^{BE}_x,
\end{equation}
and the quantum mutual information quantities $I(X;B)_\omega$ and $I(X;E)_\omega$ are the following Holevo information quantities
\begin{align}
I(X;B)_\omega&=H(\rho^B)-\int_x\!dx\,p(x)H\left(\rho^B_x\right),\\
I(X;E)_\omega&=H(\rho^E)-\int_x\!dx\,p(x)H\left(\rho^E_x\right),
\end{align}
where $H(\cdot)$ denotes the von Neumann entropy.
We also have
\begin{align}
\rho_x^B&=\sum_{|e\rangle} \langle e|\rho_x^{BE}|e\rangle,\\
\rho_x^E&=\sum_{|b\rangle} \langle b|\rho_x^{BE}|b\rangle,\\
\rho^B&=\int_x\!dx\,p(x)\rho_x^B,\\
\rho^E&=\int_x\!dx\,p(x)\rho_x^E,
\end{align}
where $\rho_x^{BE}$ is the density matrix of system $BE$ conditioned on the measurement result $X=x$. $|b\rangle$ and $|e\rangle$ each represents a complete (usually orthonormal) basis of system $B$ and $E$. Since each conditional state $|x\rangle\langle x|^X\otimes\rho_x^{BE}$ is a pure state, we have $H\left(\rho_x^B\right)=H\left(\rho_x^E\right)$. This leads to
\begin{align}
K_\rightarrow &\geq I(X; B)_\omega-I(X;E)_\omega\label{DRHolevoDiff}\\
&=H(\rho^B)-\int_x\!dx\,p(x)H\left(\rho_x^B\right)\nonumber\\
&-\left(H(\rho^E)-\int_x\!dx\,p(x)H\left(\rho_x^E\right)\right)\\
&=H(\rho^B)-H(\rho^E)\\
&=H(B)_\omega-H(E)_\omega\\
&=H(B)_\psi-H(E)_\psi\label{marginaleqdire}\\
&=H(B)_\psi-H(AB)_\psi\label{scheqdir}\\
&=I(A\rangle B)_\rho,
\end{align}
which is the expression for coherent information~\cite{devetak2005distillation, SN96} of $\rho^{AB}$. Here Eq.~(\ref{marginaleqdire}) follows from the fact that the marginal states of systems $B$ and $E$ for the states $\omega$ and $\psi$ are the same. Equation~(\ref{scheqdir}) follows from the fact that for a tripartite pure state $\psi^{ABE}$, $H(AB)_\psi=H(E)_\psi$.

For reverse reconciliation, similarly, by changing the roles of Alice and Bob, we arrive at an expression for a Hashing lower bound on the secret key distillation rate $K_\leftarrow$ given by
\begin{equation}
K_\leftarrow \geq I(A;Y)_\omega-I(Y;E)_\omega.
\end{equation}
Here the state $\omega^{AYE}$ is
\begin{equation}
\omega^{AYE}=\int_y\!dy\,p(y)|y\rangle\langle y|^Y\otimes\rho^{AE}_y,
\end{equation}
and $Y$ is the classical outcome of measuring Bob's quantum system $B$.  Now if $\psi^{ABE}$ is a pure state, $\rho_y^{AE}$ is also a pure state, which gives us $H(\rho_y^A)=H(\rho_y^E)$. And similar to the direct reconciliation case this leads to
\begin{align}
K_\leftarrow &\geq I(A;Y)_\omega-I(Y;E)_\omega\\
&=H(\rho^A)-\int_y\!dy\,p(y)H\left(\rho_y^A\right)\nonumber\\
&-\left(H(\rho^E)-\int_y\!dy\,p(y)H\left(\rho_y^E\right)\right)\\
&=H(\rho^A)-H(\rho^E)\\
&=H(A)_\omega-H(E)_\omega\\
&=H(A)_\psi-H(E)_\psi\label{marginaleqrev}\\
&=H(A)_\psi-H(AB)_\psi\label{scheqrev}\\
&=I_R(A\rangle B)_\rho,
\end{align}
which is the expression for reverse coherent information~\cite{garcia2009reverse, devetak2005distillation,horodecki2000unified} of $\rho^{AB}$. Similarly here Eq.~(\ref{marginaleqrev}) follows from the fact that the marginal states of systems $A$ and $E$ for the states $\omega$ and $\psi$ are the same. Equation~(\ref{scheqrev}) follows from the fact that for a tripartite pure state $\psi^{ABE}$, $H(AB)_\psi=H(E)_\psi$. And we also have
\begin{align}
\rho_y^A&=\sum_{|e\rangle} \langle e|\rho_y^{AE}|e\rangle,\\
\rho_y^E&=\sum_{|a\rangle} \langle a|\rho_y^{AE}|a\rangle,\\
\rho^A&=\int_y\!dy\,p(y)\rho_y^A,\\
\rho^E&=\int_y\!dy\,p(y)\rho_y^E.
\end{align}

\subsection{Restricted Eve}\label{LBrestrictedEve}
When Eve only has restricted access to the wiretapped light, i.e., $\kappa< 1$ in Fig.~\ref{fig1}, she does not have access to the full purification of the bipartite state $\rho^{AB}$ shared between Alice and Bob. That is, the systems $A$, $B$, and $E$ are not in a pure state anymore. It is rather together with the system $F$, which is lost to the environment, that these systems are in a pure state $|\psi\rangle^{ABEF}$.

In the direct reconciliation case after Alice measures system $A$ into a classical register $X$, with the tripartite state between Alice, Bob and Eve being $\rho^{ABE}=\int_x\!dx\,p(x)|x\rangle\langle x|^A\otimes\rho_x^{BE}$, we have
\begin{align}
K_\rightarrow &\geq I(X;B)_\omega-I(X;E)_\omega\\
&=H(\rho^B)-H(\rho^E)-\int_x\!dx\,p(x)\left(H\left(\rho_x^B\right)-H\left(\rho_x^E\right)\right)\label{DireLBeq}.
\end{align}
The second term $\int_x\!dx\,p(x)\left(H\left(\rho_x^B\right)-H\left(\rho_x^E\right)\right)$ now does not vanish. Similarly, in reverse reconciliation case we have
\begin{align}
K_\leftarrow &\geq I(A;Y)_\omega-I(Y;E)_\omega\\
&=H(\rho^A)-H(\rho^E)-\int_y\!dy\,p(y)\left(H\left(\rho_y^A\right)-H\left(\rho_y^E\right)\right)\label{ReveLBeq}.
\end{align}

\section{Upper Bound for Secret Key Distillation under Restricted Eavesdropping}\label{UBs}


In this Section, we recall the relative entropy of entanglement of a channel~\cite{pirandola2017fundamental} ($E_{\textrm{R}}$), which serves as an upper bound on the entanglement and secret key distillation capacities of the channel under unrestricted eavesdropping when assisted by unlimited two-way classical communication assistance between the communicating parties, and apply it to the restricted eavesdropping model. (See also~\cite{tomamichel2017strong} for a related definition
of Rains information of a channel, which is relevant
specifically in the context of entanglement distillation.)

\begin{definition}
	The relative entropy of entanglement of a channel $\mathcal{N}_{A'\rightarrow B}$ is defined as~\cite{pirandola2017fundamental}
	\begin{equation}
	E_{\textrm{R}}(\mathcal{N}):=\sup_{\phi^{AA'}}E_{\operatorname{R}}\left(A;B\right)_{\rho},
	\end{equation}
	where
	\begin{equation}
	E_\textrm{R}(\rho):=\inf_{\sigma\in \textrm{SEP}}D(\rho||\sigma),\label{REEDef}
	\end{equation}
	$\rho^{AB}=\mathcal{N}_{A'\rightarrow B}\left(\phi^{AA'}\right)$, and
	$D(\rho||\sigma)$ is the relative entropy between states $\rho$ and $\sigma$. When the support of $\rho$ contains that of $\sigma$,
	\begin{equation}
	D(\rho||\sigma):=\operatorname{Tr}\left(\rho\left(\log\rho-\log\sigma\right)\right).\label{DivDef}
	\end{equation}
	While the relative entropy of entanglement is the relative entropy of a state with its closest separable (SEP) state in Hilbert space, the relative entropy of entanglement of a channel is the relative entropy of entanglement of the state distributed across the channel optimized over all possible inputs to the channel. 
	\label{reechannel}
\end{definition}

\subsubsection{Unrestricted Eve}
Using the relative entropy of entanglement, Pirandola \emph{et al}. (PLOB)~\cite{pirandola2017fundamental} gave an upper bound to the energy-unconstrained, two-way unlimited Local Operations and Classical Communication (LOCC)-assisted
entanglement and secret key distillation capacity of lossy and noisy bosonic channels. For a pure loss channel $\mathcal{N}_\eta$ of transmissivity $\eta$, the relative entropy of entanglement upper bound is given by $\max\{K_\rightarrow, K_\leftarrow\}\leq E_\textrm{R}(\mathcal{N}_\eta)=-\log_2(1-\eta)$. Since this upper bound matches the reverse coherent information lower bound~\cite{garcia2009reverse,pirandola2009direct}, the above rate characterizes the capacity. (See also \cite{WTB17} for a strong converse theorem for the upper bound.)

For a thermal noise channel $\mathcal{N}_{\eta,n_e}$ of transmissivity $\eta$ and thermal noise $n_e$, which is the mean photon number in the thermal state that Eve injects into the channel, the relative entropy of entanglement upper bound is known to be~\cite{pirandola2017fundamental} $\max\{K_\rightarrow, K_\leftarrow\}\leq E_\textrm{R}(\mathcal{N}_{\eta,n_e})=\log_2\left(\frac{\eta}{1-\eta}\right)-g\left(n_e\right)$, where~\cite{Serafini17,weedbrook2012gaussian} 
\begin{align}
g(x)=(x+1)\log_2(x+1)-x\log_2(x)\label{thentropy}
\end{align}
is the von Neumann entropy of a thermal  state of mean photon number $x$.

The bipartite separable quantum states closest in relative entropy divergence to a two-mode entangled quantum states shared across the channel that are required in giving the above $E_{\textrm{R}}$ bounds are found using the Positive-Partial Transpose (PPT) criterion for separability of quantum states~\cite{HORODECKI96, Peres96}.

\subsubsection{Restricted Eve}
In the restricted eavesdropping model in Fig.~\ref{fig1}, the state of interest now is the tripartite state $\rho^{AFB}$, which is purified by the fourth system $E$ that Eve possesses (unlike the unrestricted case, where system $F$ does not exist and the state of interest is bipartite). We apply the PPT criterion across the bipartition $AF$ and $B$ and give relevant $E_{\textrm{R}}$ upper bounds in this scenario. See Appendix A for details of the calculation.

\section{Achievable Rate and Upper Bound Derivation with Numerical Results}\label{results}
In this Section, we apply the methods of secure key rate (SKR) analysis and upper bounds presented in Secs.~\ref{LBs} and \ref{UBs} to bosonic pure loss and thermal noise channels fed with an input two-mode squeezed vacuum (TMSV) state $|\Psi\rangle^{AA'}=(\cosh r)^{-1} \sum_{n=0}^\infty (\tanh r)^n|n\rangle|n\rangle$. The achievable rates are given for heterodyne detection either at Alice or Bob, which correspond to direct and reverse information reconciliation scenarios, respectively.

\subsection{Achievable Rates}

\subsubsection{Pure Loss Channel}

\begin{figure}[H] 
\centering{\includegraphics[width=0.5\textwidth]{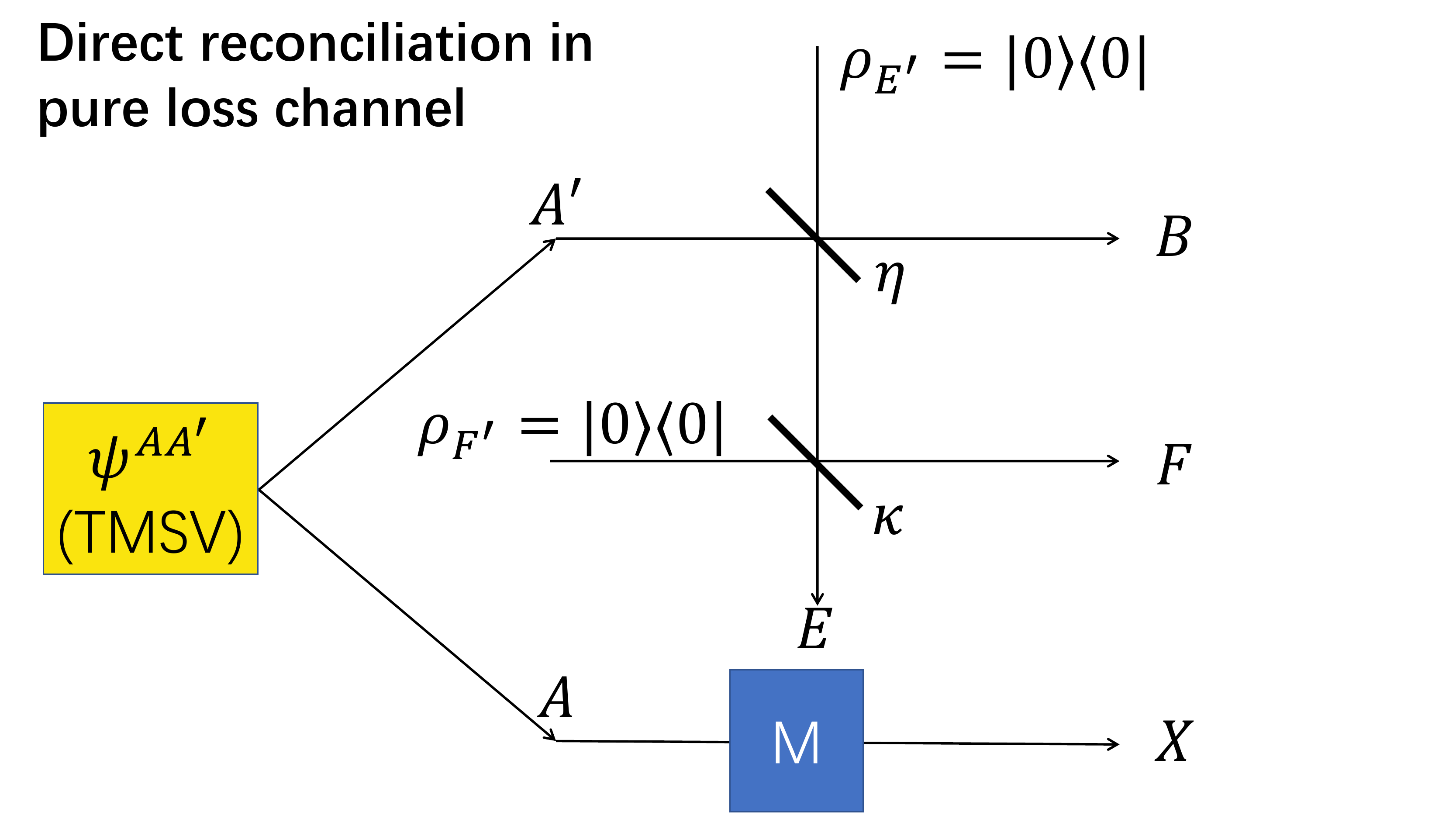}}
\caption{Entanglement-based model for secret key distillation over a pure loss bosonic channel based on heterodyne detection and direct reconciliation. Here Alice performs heterodyne measurement of system $A$ and this projects the system $A'$ of the TMSV state $\psi^{AA'}$ (with $\mu$ mean photon number per mode) onto a coherent state $|\alpha\rangle^{A'}$. She then sends side information in the classical channel to Bob to help him distill keys from his system. Here vacuum states are injected from $E'$ and $F'$ denoting a pure loss channel. The restriction on Eve is imposed by letting only a $\kappa$ fraction of the wiretapped light reach her receiver. \label{DirePLPic}}
\end{figure}

First we will show the achievable rate with direct reconciliation, namely where Alice performs heterodyne detection on her system, as depicted in Fig.~\ref{DirePLPic}. Assuming a TMSV state input, we calculate the achievable rate for this setup.

Since the heterodyne measurement on $A$  projects the other part $A'$ of the TMSV onto a coherent state $\rho_x^{A'}=|\alpha\rangle$, we know that the state at the beam splitters' outputs conditioned on measurement result $x$, namely $\rho_x^B$, $\rho_x^E$, $\rho_x^F$ are also coherent states with attenuated amplitudes. Since they are pure states we have
\begin{equation}
H\left(\rho_x^B\right)=H\left(\rho_x^E\right)=0.
\end{equation}
\noindent
So, using Eq.~(\ref{DireLBeq}), we have
\begin{align}
K_\rightarrow &\geq H(\rho^B)-H(\rho^E)\\
&=g\left(\eta\mu\right)-g\left(\kappa\mu\left(1-\eta\right)\right),\label{DirePLReq1}
\end{align}
where $\mu=\sinh^2(r)$ is the average photon number in the light Alice transmits to Bob, which leads to
\begin{equation}
\centering
\lim_{\mu\rightarrow \infty}K_\rightarrow \ge \log_2\frac{\eta}{\kappa(1-\eta)}.\label{DirePLReq1L}
\end{equation}
\noindent
Equation~(\ref{DirePLReq1L})~\cite{holevo2001evaluating} gives the limiting value of the key rate when the input photon number is taken to infinity. This limit can be shown to be the optimal input strength that maximizes the key rate. Notice that the dependence of the direct reconciliation achievable rate in Eq.~(\ref{DirePLReq1L}) on Eve's restriction $\kappa$ is in the denominator inside the log function. Thus, restricting Eve's received power can help increase the achievable rate beyond the rate achievable against an unrestricted Eve ($\kappa=1$ in Eq.~(\ref{DirePLReq1L})), namely $\log_2\left(\frac{\eta}{1-\eta}\right)$. It is interesting to note that the increase in achievable rate is accomplished without affecting the channel from Alice to Bob, but rather by modifying the channel from Alice to Eve.

In the case of an unrestricted Eve and direct reconciliation, we need to have $\eta>(1-\eta)$ to attain a positive key rate in Eq.~(\ref{DirePLReq1L}). This gives us $\eta>0.5=3\textrm{dB}$ which is known as the "3dB limit" for direct reconciliation where key rate drops to zero when channel transmissivity is below 3dB. Similarly for the key rate to be greater than zero in the restricted Eve case, we need to have $\eta>\kappa(1-\eta)$, which gives $\eta>\frac{\kappa}{1+\kappa}$. This condition captures the limitation of direct reconciliation with regard to the transmission distance, namely that the key rate vanishes beyond a threshold distance.

\begin{figure}[htbp] 
\centering{\includegraphics[height=12pc]{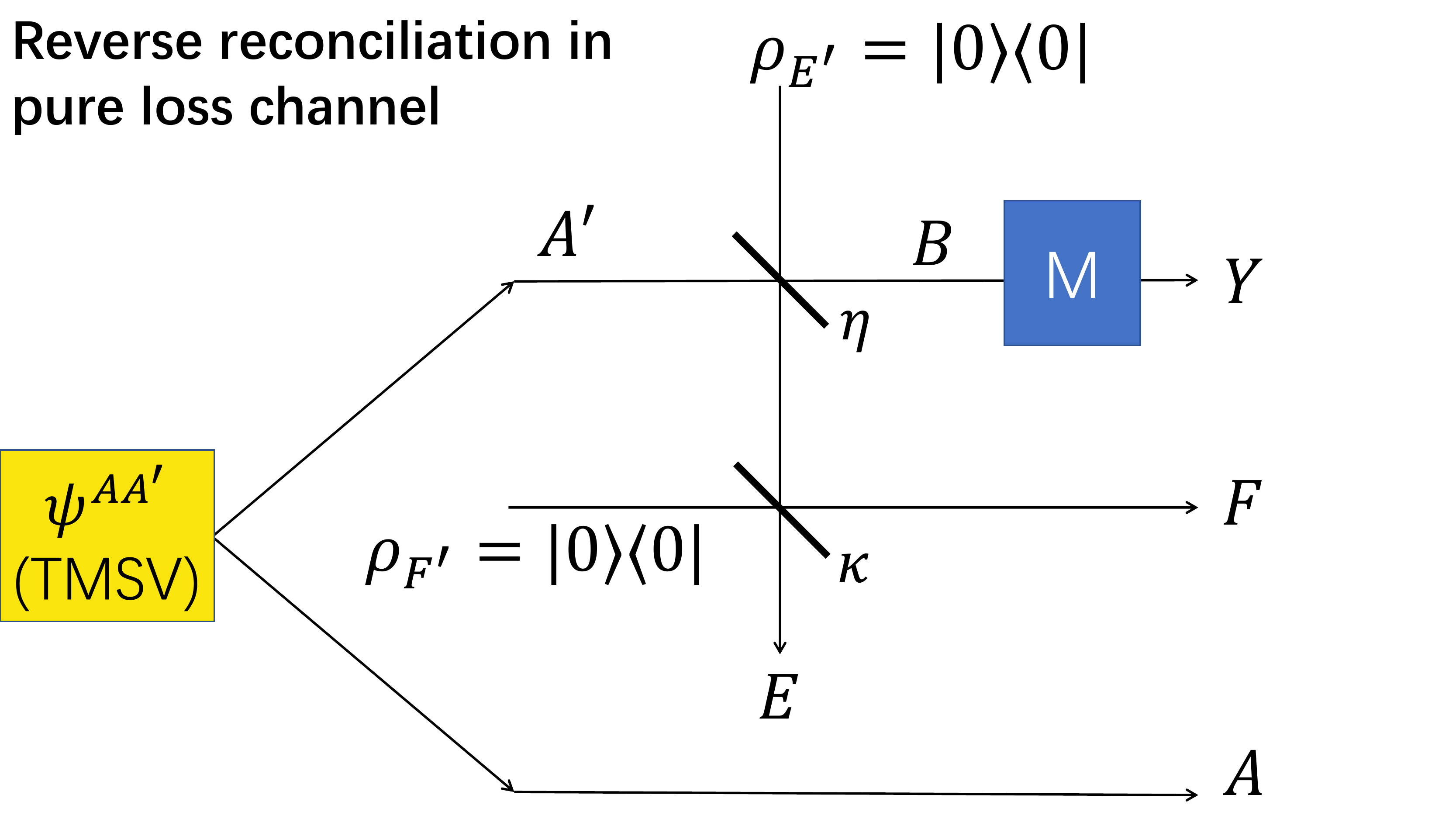}}
\caption{Entanglement-based model for secret key distillation over a pure loss bosonic channel based on heterodyne detection and reverse reconciliation. Here Bob performs heterodyne measurement on his system $B$; the  states injected from $E'$ and $F'$ are vacuum states. \label{RevePLPic}}
\end{figure}

Now, consider the case of reverse reconciliation, as depicted in Fig.~\ref{RevePLPic}. Here, Bob performs heterodyne measurement on his system $B$ and sends side information through a classical communication channel to Alice to help her distill secret key. Using Eq.~(\ref{ReveLBeq}), and recognizing that $Y$ is a continuous variable, we  get

\begin{align}
K_\leftarrow &\geq H(\rho^A)-H(\rho^E)-\int_y\!dy\,p(y)\left(H\left(\rho_y^A\right)-H\left(\rho_y^E\right)\right)\\
&=g(\mu)-g(\kappa\mu(1-\eta))\nonumber\\
&-\int_y\!dy\,p(y)\left(g\left(\frac{\mu(1-\eta)}{1+\eta\mu}\right)-g\left(\frac{(1-\eta)\kappa\mu}{1+\eta\mu}\right)\right)\label{RevePLReq2}\\
&=g(\mu)-g(\kappa\mu(1-\eta))\nonumber\\
&-\left(g\left(\frac{\mu(1-\eta)}{1+\eta\mu}\right)-g\left(\frac{(1-\eta)\kappa\mu}{1+\eta\mu}\right)\right)\label{RevePLReq4},
\end{align}
and
\begin{align}
\lim_{\mu\rightarrow \infty}K_\leftarrow &\ge \log_2\frac{1}{\kappa(1-\eta)}-\left(g\left(\frac{1-\eta}{\eta}\right)-g\left(\frac{(1-\eta)\kappa}{\eta}\right)\right)\label{RevePLReqL}.
\end{align}

Since in this case  the post-measurement conditional states are not pure, we derived the covariance matrix of corresponding states and calculated the von Neumann entropies, which turns out as shown in Eq.~(\ref{RevePLReq2}). Since the argument of the $h$ functions in Eq.~(\ref{RevePLReq2}) have no dependence on the probability distribution of $y$, we get Eq.~(\ref{RevePLReq4}) by integrating over $y$. Taking the limit of input mean photon number $\mu\rightarrow\infty$,  we obtain the optimal achievable rate given in Eq.~(\ref{RevePLReqL}).

Equation~(\ref{RevePLReqL}), when $\kappa=1$, reduces to $-\log_2(1-\eta)$, which is also the PLOB bound based on $E_\textrm{R}$, discussed in Sec.~\ref{UpBR}, and hence the capacity. If we compare Eq.~(\ref{RevePLReqL}) for $\kappa<1$ with pure loss unrestricted model capacity $-\log_2(1-\eta)$, not only do we have $\kappa$ showing up in the denominator inside the logarithm, but we also have one correction term: $-(g(\frac{\mu(1-\eta)}{1+\eta\mu})-g(\frac{(1-\eta)\kappa\mu}{1+\eta\mu}))$. This correction term changes differently with $\kappa$ compared to the first term $\log_2(\frac{1}{\kappa(1-\eta)})$. Unlike the case with the unrestricted Eve, we find that the achievable rate with reverse reconciliation is not always better than the rate with direct reconciliation. 

\begin{figure}[H] 
\centering
{\includegraphics[height=17pc]{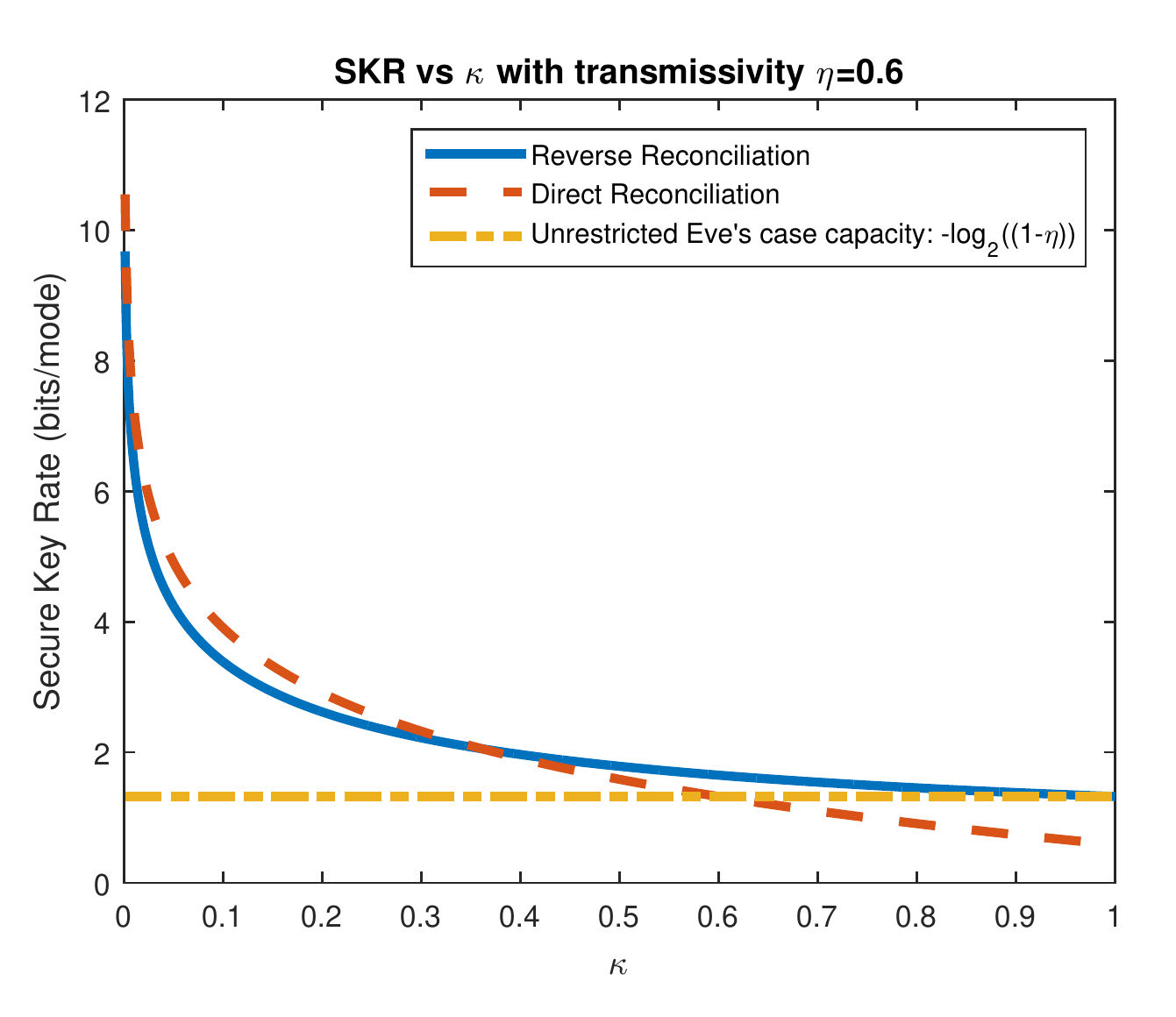}\\
\includegraphics[height=0.8pc]{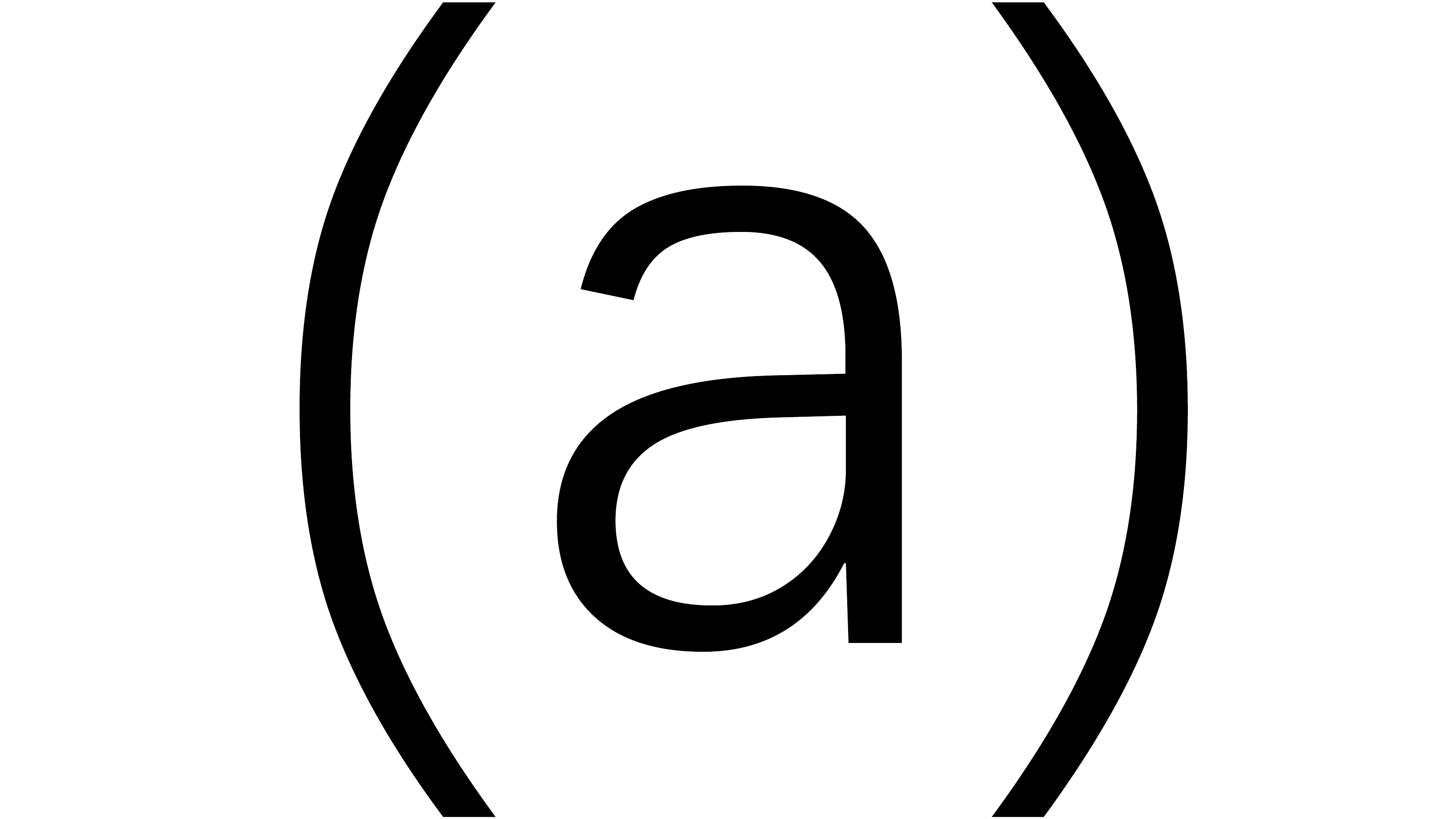}}\\
{\includegraphics[height=18pc]{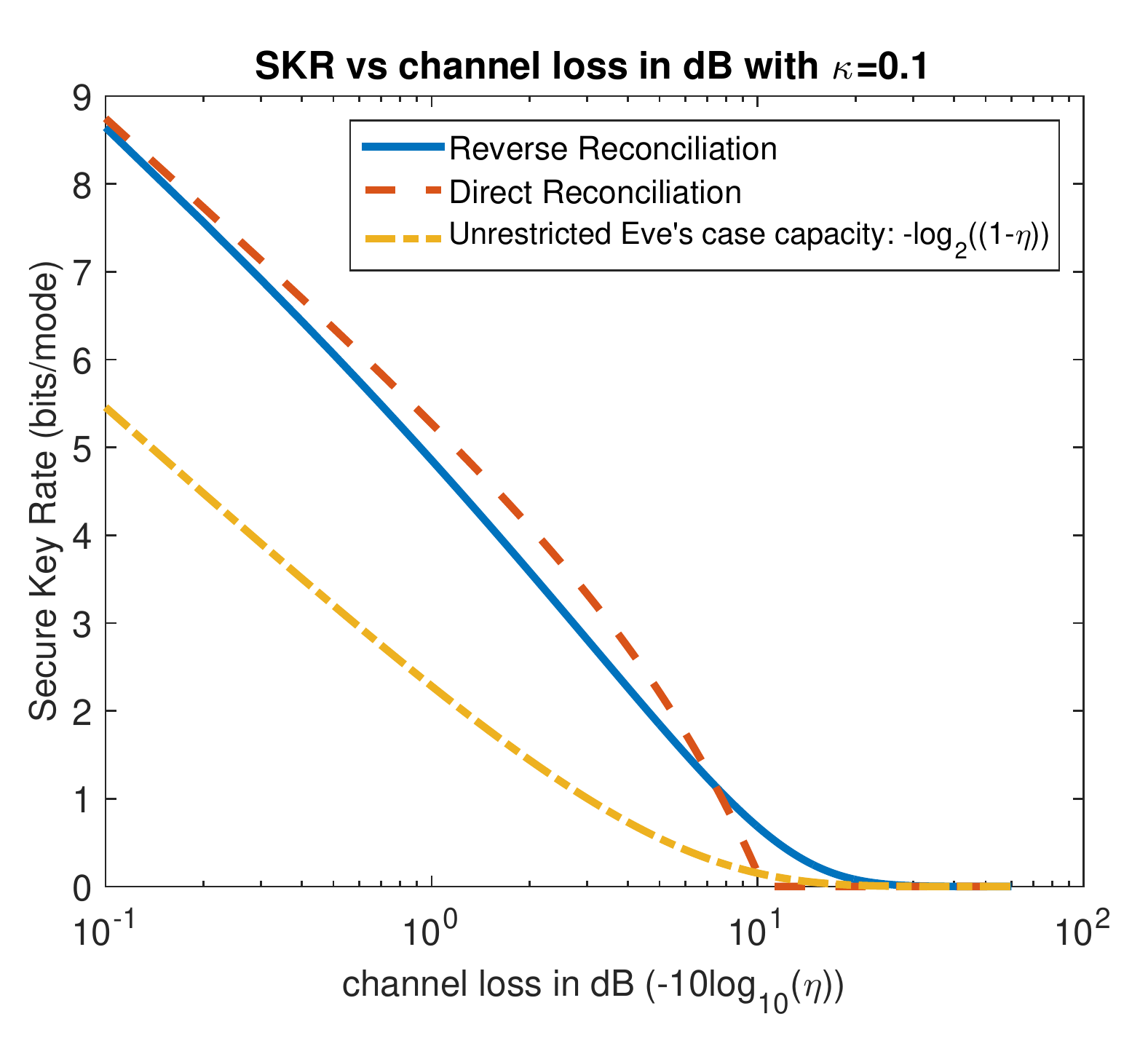}\\
\includegraphics[height=0.8pc]{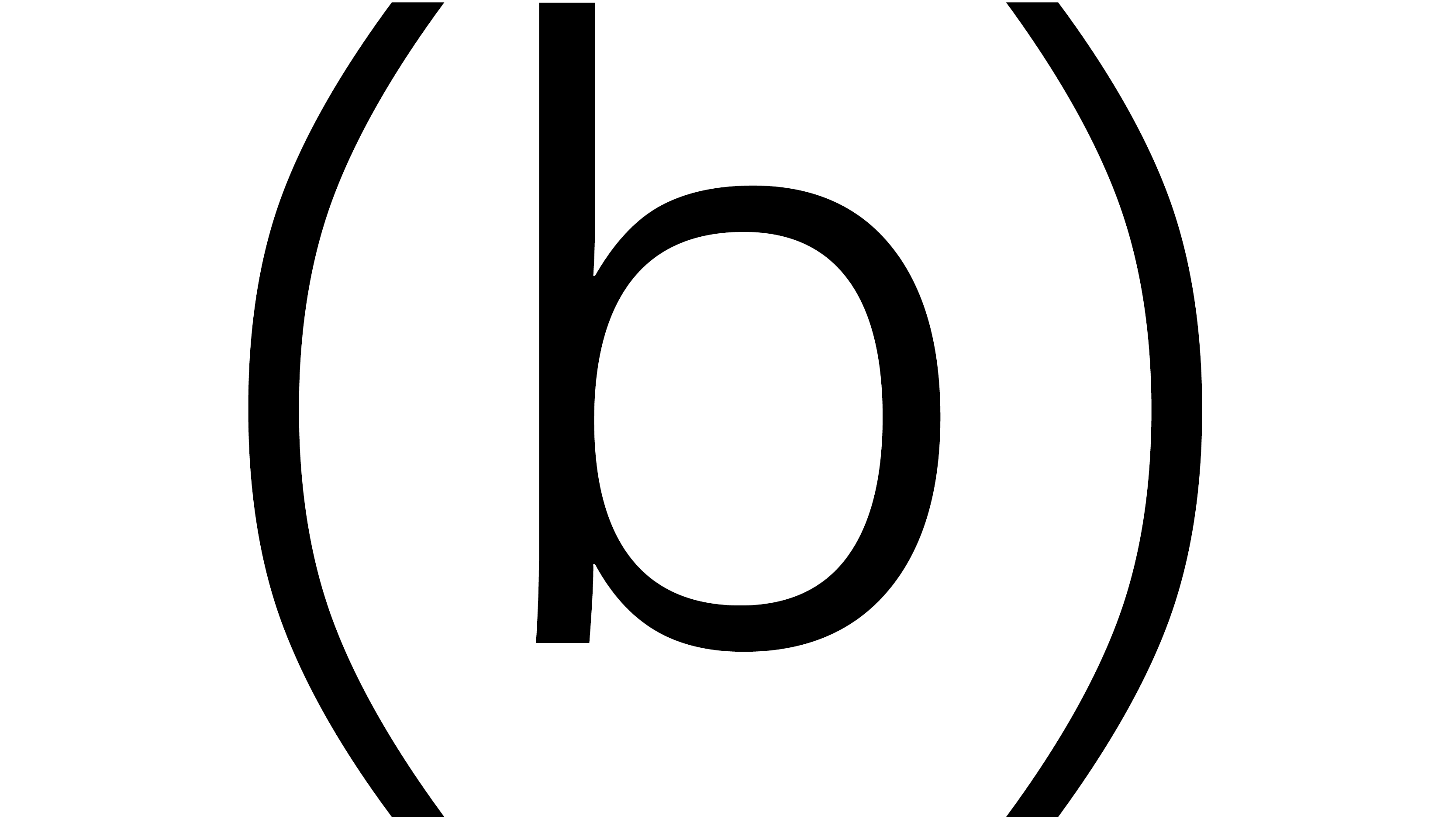}}
\caption{(a) Secure key rate against restricted eavesdropping over a pure loss channel with TMSV state and heterodyne detection. Direct reconciliation vs. reverse reconciliation  rates as a function of $\kappa$. Here the channel transmissivity is set to $\eta=0.6$. (b) Direct reconciliation vs. reverse reconciliation achievable rate as a function of channel loss in dB  with $\kappa=0.1$. The channel capacity against an unrestricted Eve is also shown for comparison.\label{plSKRvKorL}}
\end{figure}

In Fig.~\ref{plSKRvKorL} (a), we plot the SKR as a function of the restriction on Eve $0<\kappa<1$ for a channel transmissivity of $\eta=0.6$. When high values of $\kappa$ are assumed, which includes the unrestricted Eve's case ($\kappa=1$), we find that reverse reconciliation gives a higher achievable rate than direct reconciliation. However  when low values of $\kappa$ are assumed the rate with direct reconciliation is seen to exceed the rate with reverse reconciliation.

In Fig.~\ref{plSKRvKorL} (b), we plot the direct and reverse reconciliation achievable rates as a function of the channel transmissivity $\eta$ for a given value of the restriction on Eve $\kappa=0.1$. Since channel loss usually increases as transmission distance increases, we can see that the  reverse reconciliation scheme has a longer useful transmission distance than the direct reconciliation scheme, which is similar to the case when Eve is unrestricted as was shown in ~\cite{garcia2009reverse}. However, here both direct reconciliation and reverse reconciliation can achieve rates that are higher than the unrestricted Eve's capacity, which was achievable with reverse reconciliation.

\subsubsection{Thermal Noise Channel}

For the thermal noise channel, Eve's input to the channel is now a thermal state of mean photon number $n_e$. 

Since the thermal state is not a pure state, we need to take into account its purifying system $R$. So, here we assume that Eve holds a TMSV with mean photon number $n_e$. She injects one mode $E'$ into the channel while keep the other mode as a purification and then does a joint operation on systems $E$ and $R$.

\begin{figure}[H] 
\centering{\includegraphics[height=12pc]{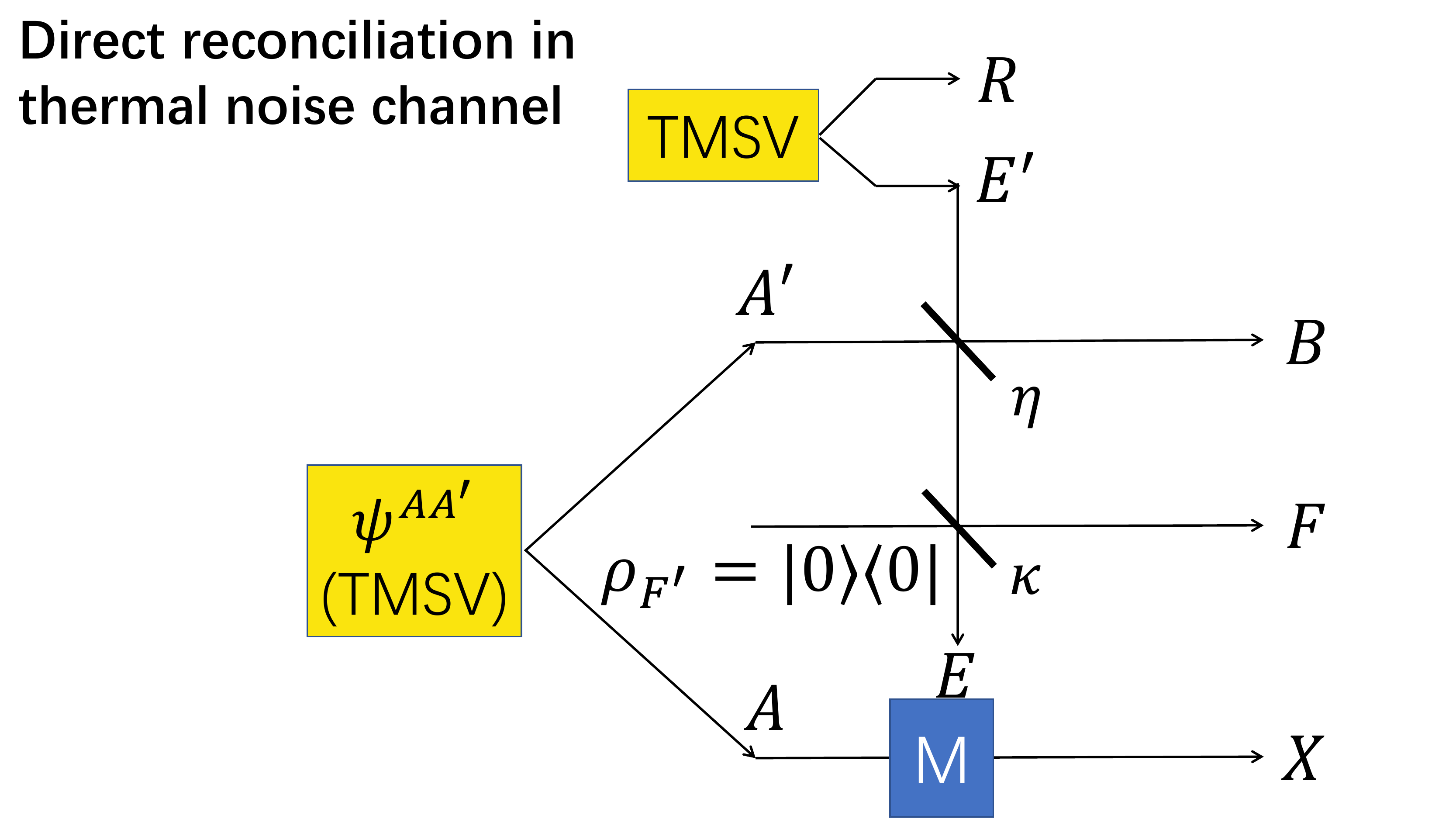}}
\caption{Entanglement-based model for secret key distillation over a thermal noise bosonic channel based on heterodyne detection and direct reconciliation. Here Alice performs heterodyne measurement and sends side information to Bob to help him distill secret keys. Eve holds a TMSV with $n_e$ mean photon number per mode and injects one of the modes into the channel, actively inducing noise.\label{Dnoisy}}
\end{figure}

First we look at the case of direct reconciliation, depicted in Fig.~\ref{Dnoisy}. Again using Eq.~(\ref{DireLBeq}), we have
\begin{align}
K_\rightarrow &\geq I(X;B)_\omega-I(X;ER)_\omega\\
&=H(\rho^B)-H(\rho^{ER})-\int_x\!dx\,p(x)\left(H\left(\rho_x^B\right)-H\left(\rho_x^{ER}\right)\right)\\
&=g(n_e(1-\eta)+\eta\mu)-\sum_i g\left(\frac{\nu^{ER}_i-1}{2}\right)\nonumber\\
&-(g(n_e(1-\eta))-(g(n_e(1-\eta\kappa)).\label{eq:3}
\end{align}
Here the state $\omega^{XBER}$ is
\begin{equation}
\omega^{XBER}=\int_x\!dx\,p(x)|x\rangle\langle x|^X\otimes\rho^{BER}_x.
\end{equation}
$\nu^{ER}_i$s are the symplectic eigenvalues of the reduced covariance matrix of modes $E$ and $R$. Finally, we have
\begin{align}
\lim_{\mu\rightarrow\infty}K_\rightarrow&\ge \log\frac{\eta}{\kappa(1-\eta)}-g(n_e)\nonumber\\
&-g(n_e(1-\eta))+g(n_e(1-\eta\kappa))\label{DireLBHL}.
\end{align}

Here if we compare Eq.~(\ref{DireLBHL}) with the direct reconciliation achievable rate for an unrestricted Eve case over the thermal noise channel~\cite{pirandola2017fundamental} $\log(\frac{\eta}{(1-\eta)})-g(n_e)$, we can see that $\kappa$ shows up in the denominator inside the logarithm function in a way similar to the case of the pure loss channel. However, here we still have a correction term $-g(n_e(1-\eta))+g(n_e(1-\eta\kappa))$, which vanishes when $\kappa\rightarrow 1$.

\begin{figure}[H] 
\centering{\includegraphics[height=12pc]{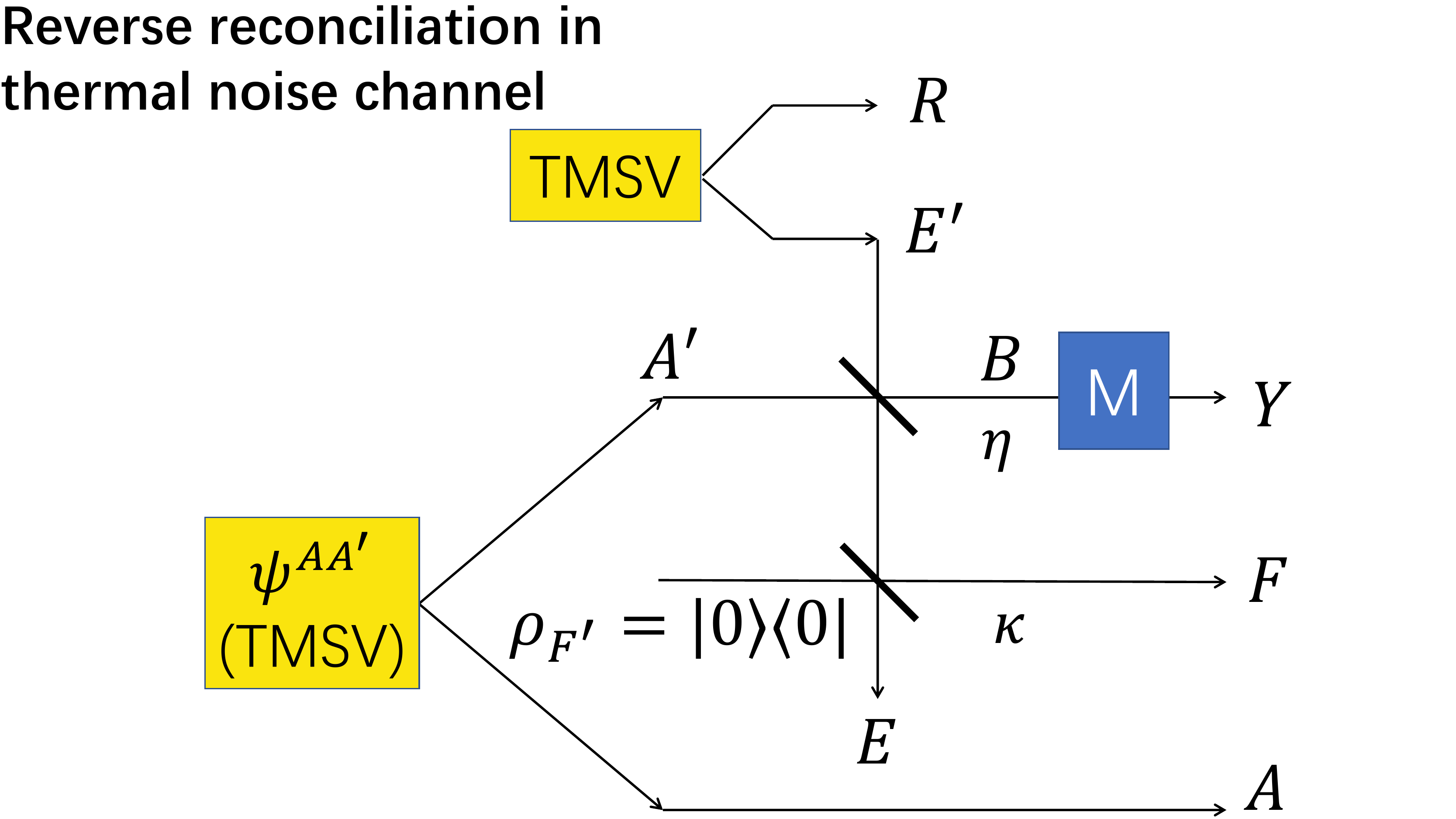}}
\caption{Entanglement-based model for secret key distillation over a thermal noise bosonic channel based on heterodyne detection and reverse reconciliation. Here Bob performs heterodyne measurement and sends side information to Alice to help her distill secret keys. Eve holds a TMSV with $n_e$ mean photon number per mode and injects one of the modes into the channel, actively inducing noise.\label{Rnoisy}}
\end{figure}

Now, for the reverse reconciliation case depicted in Fig.~\ref{Rnoisy}, we use Eq.~(\ref{ReveLBeq}) to obtain our results
\begin{align}
K_\leftarrow &\geq I(A;Y)_\omega-I(Y;ER)_\omega\\
&=H(\rho^A)-H(\rho^{ER})-\int_y\!dy\,p(y)\left(H\left(\rho_y^A\right)-H\left(\rho_y^{ER}\right)\right)\\
&=g(\mu)-\sum_i g\left(\frac{\nu^{ER}_i-1}{2}\right)\nonumber\\
&-g\left(\mu-\frac{\eta\mu(1+\mu)}{1+n_e-n_e\eta+\eta\mu}\right)+\sum_i g\left(\frac{\nu^{ER}_{y_i}-1}{2}\right).
\end{align}
Here the state $\omega^{AYER}$ is
\begin{equation}
\omega^{AYER}=\int_y\!dy\,p(y)|y\rangle\langle y|^Y\otimes\rho^{AER}_y.
\end{equation}
$\nu^{ER}_i$ ($i=1,2$) are the symplectic eigenvalues of the covariance matrix corresponding to system $ER$ and $\nu^{ER}_{y_i}$ ($i=1,2$) are the symplectic eigenvalues of the covariance matrix corresponding to the post-measurement system $ER|y$ after the measurement of Bob's state $B$. Finally, we have
\begin{align}
\lim_{\mu\rightarrow\infty}K_\leftarrow&\ge \log\frac{1}{\kappa(1-\eta)}-g(n_e)-g\left(\frac{1+n_e-n_e\eta-\eta}{\eta}\right)\nonumber\\
&+\sum_i g\left(\frac{\lim_{\mu\rightarrow\infty}\nu^{ER}_{y_i}-1}{2}\right).\label{ReveLBHL2}
\end{align}
In Eq.~(\ref{ReveLBHL2}) we have $\kappa$ showing up in the denominator inside the log function compared to the reverse reconciliation achievable rate for thermal noise channels under unrestricted Eve's case $\log_2\left(\frac{1}{\kappa\left(1-\eta\right)}\right)-g\left(n_e\right)$. Also, we have correction terms of which the exact form is given below:
\begin{align}
\lim_{\mu\rightarrow\infty}&\nu^{ER}_{y_1}=\left| \frac{A-B+C+D}{\eta}\right|,\label{LB55}\\
\lim_{\mu\rightarrow\infty}&\nu^{ER}_{y_2}=\left| \frac{-A+B-C+D}{\eta}\right|,\\
A&=\eta ^2 \left(2 n_e (n_e+1)+1\right),\\
B&=2 \eta  \kappa  \left(\eta +2 n_e^2+n_e-1\right),\\
C&=2 \kappa ^2 (-\eta +n_e+1)^2,\\
D&=2 \sqrt{\left(E-F+G\right) \left(-\eta  \kappa +\kappa +n_e (\kappa -\eta )\right)^2},\\
E&=\eta ^2 (-\kappa +n_e+1)^2,\\
F&=2 \eta  \kappa  (n_e+1) (\kappa +n_e-1),\\
G&=\kappa ^2 (n_e+1)^2\label{LB63}.
\end{align}

\begin{figure}[H] 
\centering
{\includegraphics[height=11.3pc]{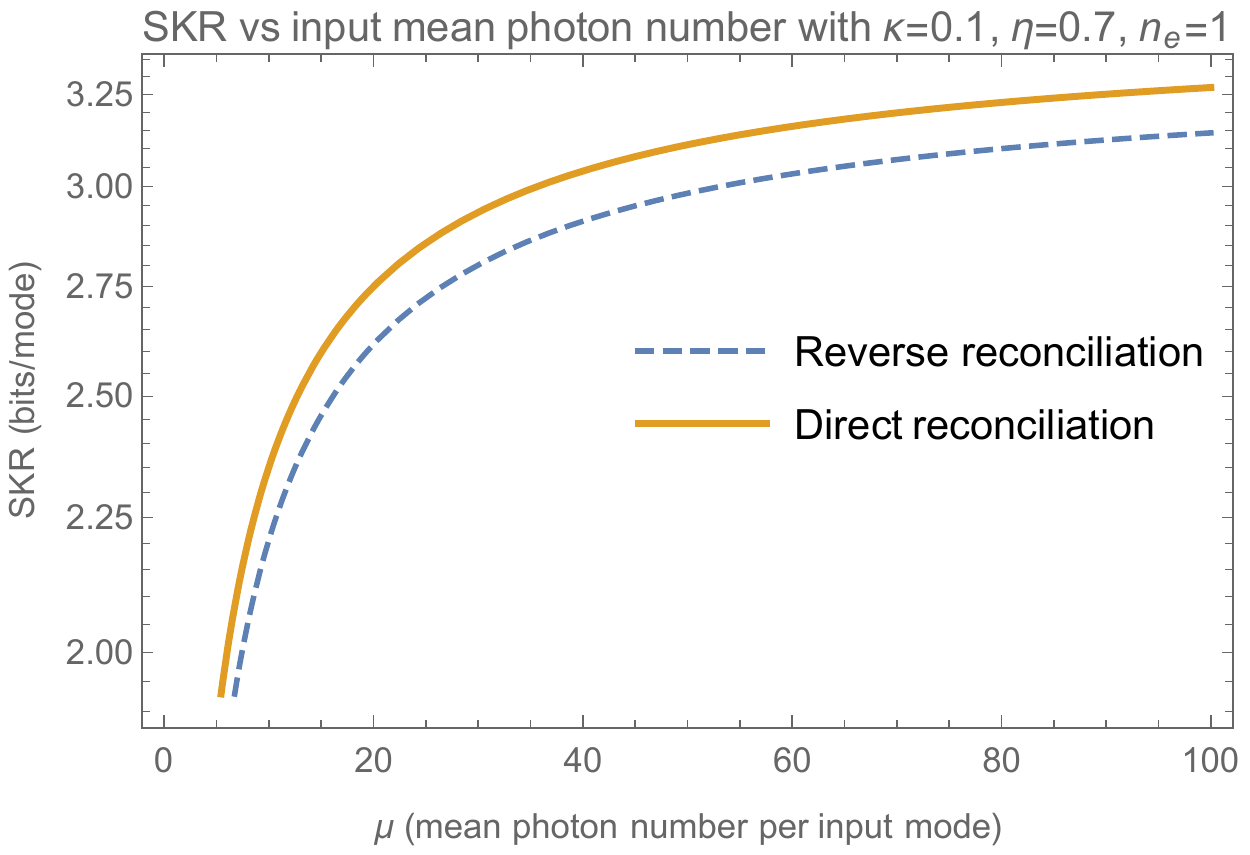}\\
\includegraphics[height=0.8pc]{a.pdf}}\\
{\includegraphics[height=11.3pc]{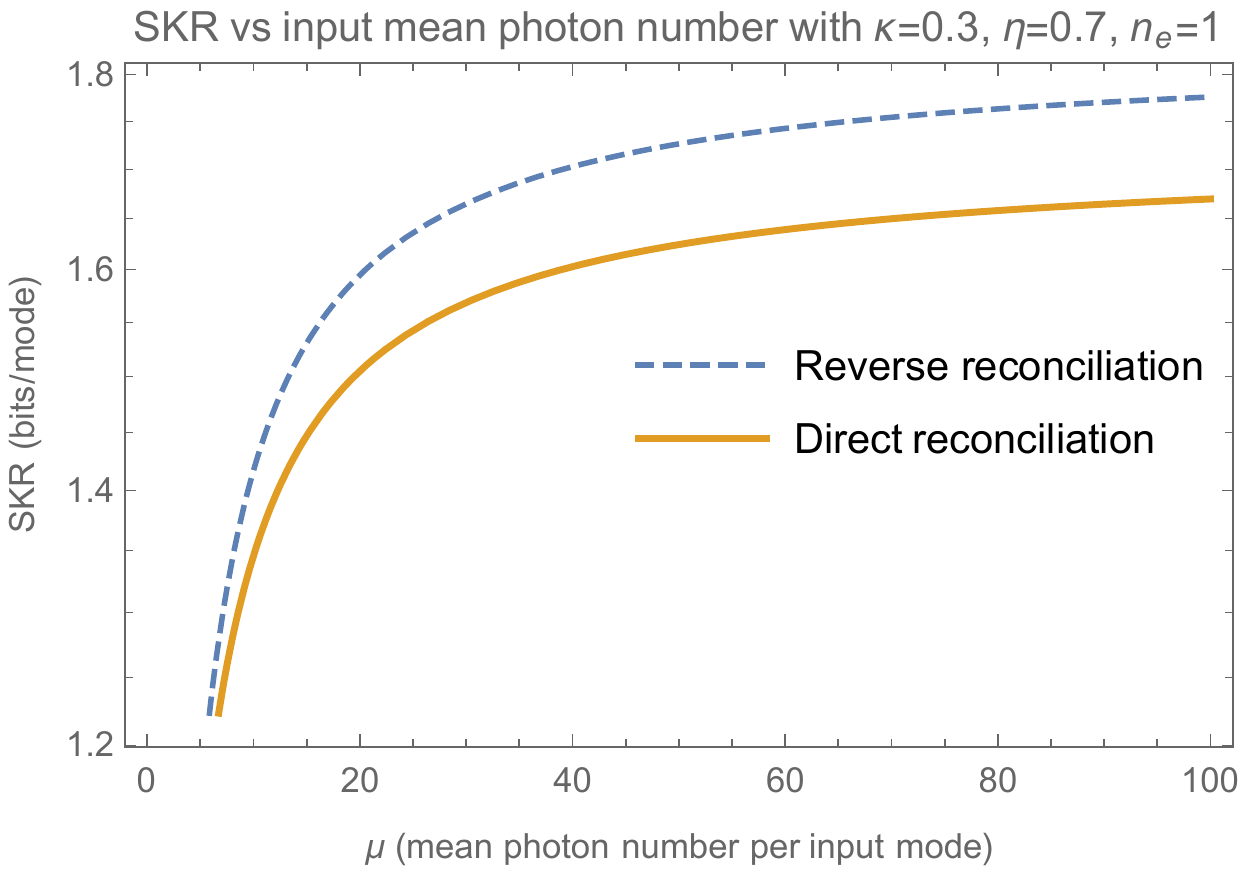}\\
\includegraphics[height=0.8pc]{b.pdf}}
\caption{Achievable rates, direct reconciliation vs. reverse reconciliation, as a function of input mean photon number $\mu$ in the input signal. With different choices of the channel parameters direct reconciliation can have a higher or lower rate than reverse reconciliation. In both figures direct and reverse SKR increase with increasing input power and saturate to some value while different $\kappa$ values change the comparison between direct and reverse reconciliation. \label{SKRvmu}}
\end{figure}

With the above results Eqs.~(\ref{LB55})-(\ref{LB63}), we first plot the direct and reverse reconciliation achievable rates as functions of the input mean photon number $\mu$ in Fig.~\ref{SKRvmu}.

Here in Fig.~\ref{SKRvmu}, we can see that the achievable rate for both direct reconciliation and reverse reconciliation is increasing with increasing input mean photon number $\mu$. So $\mu=\infty$ is  optimal. Also, depending on the channel parameters, either direct or reverse reconciliation could have the higher rate.

For the figures below, unless specified otherwise, the achievable rate is plotted with $\mu$ taken to infinity.

\begin{figure}[H] 
\centering{\includegraphics[height=12pc]{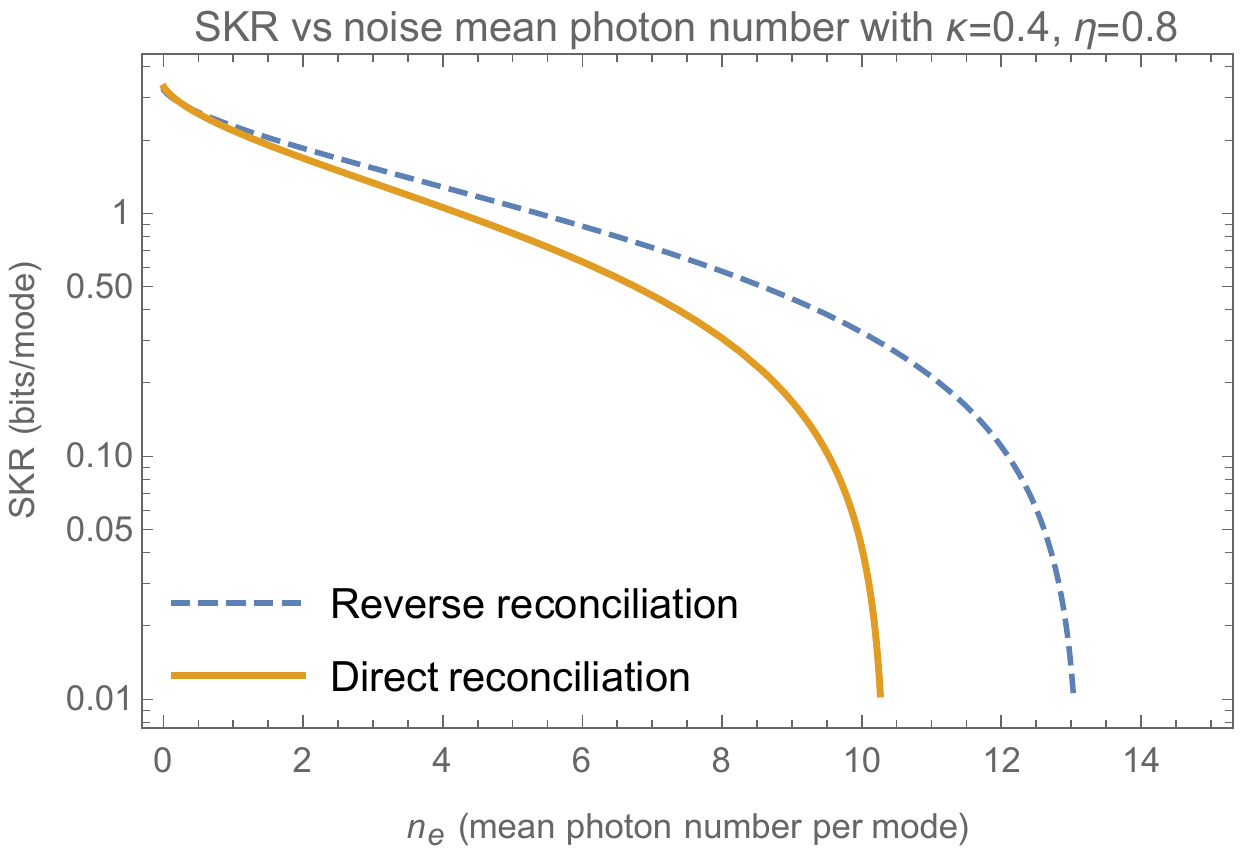}}
\caption{Achievable rate as a function of input mean photon number $n_e$ of thermal noise state with $\kappa=0.4$, $\eta=0.8$, $\mu=\infty$. \label{SKRvne}}
\end{figure}

\begin{figure}[H] 
\centering{\includegraphics[height=12pc]{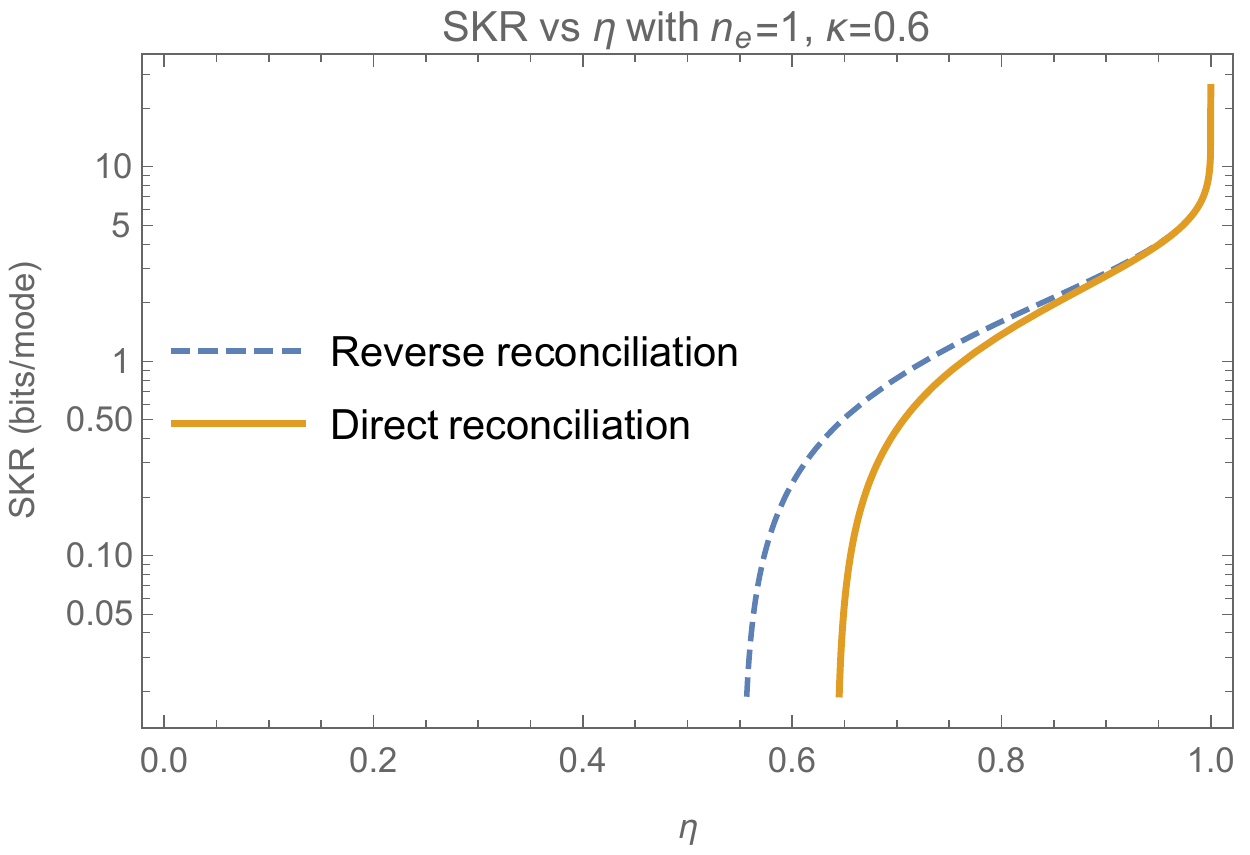}}
\caption{Achievable rate as a function of transmissivity $\eta$ with $\kappa=0.6$, $n_e=1$, $\mu=\infty$. \label{SKRveta}}
\end{figure}

\begin{figure}[H] 
\centering{\includegraphics[height=12pc]{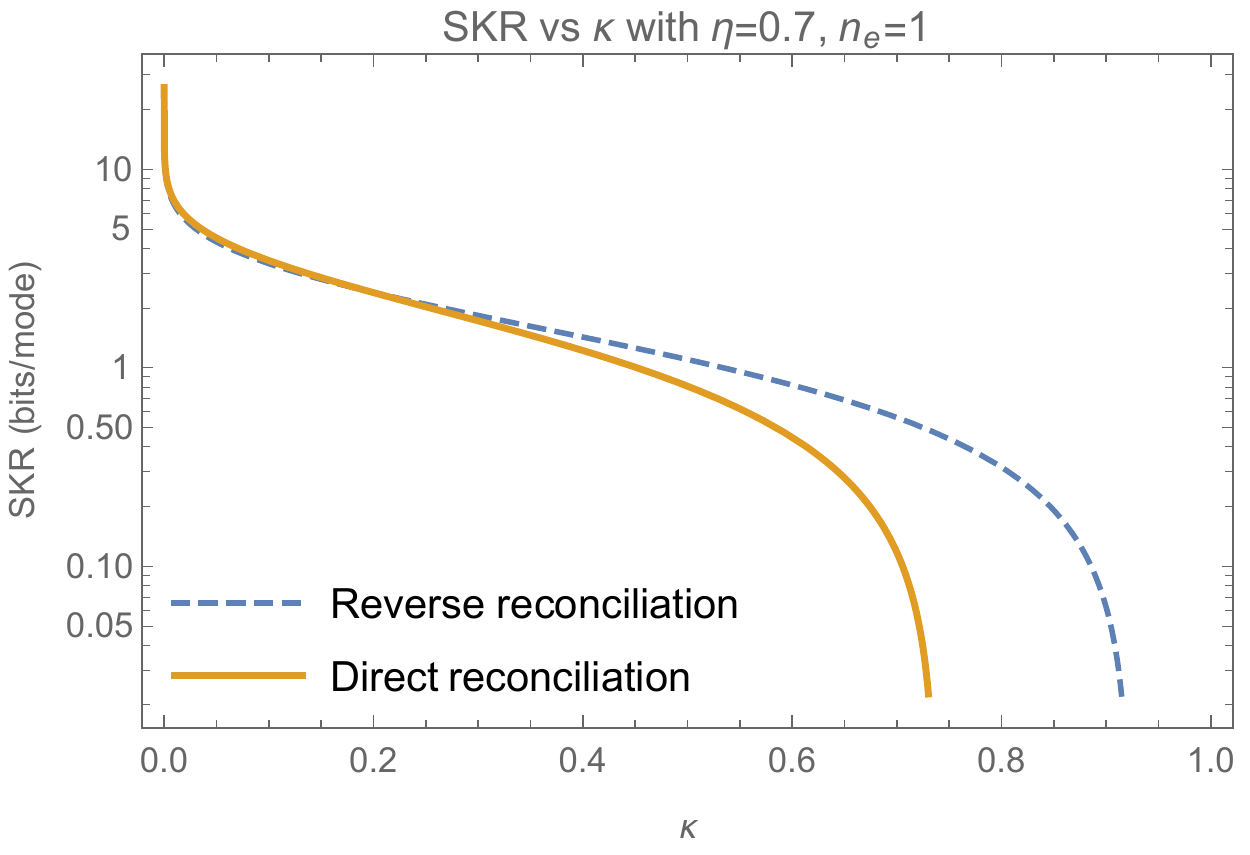}}
\caption{Achievable rate as a function of  $\kappa$ with $\eta=0.7$, $n_e=1$, $\mu=\infty$.\label{SKRvkappa}}
\end{figure}

In Figs.~\ref{SKRvne}-\ref{SKRvkappa}, we separately plot the direct and reverse reconciliation achievable rates as functions of the thermal noise strength $n_e$, channel transmissivity $\eta$, and the restriction factor $\kappa$. We can see that both direct and reverse reconciliation key rates decrease with increasing noise, increasing $\kappa$ and increasing channel loss ($-10\log_{10}(\eta)$).

\begin{figure}[H] 
\centering
{\includegraphics[height=11.8pc]{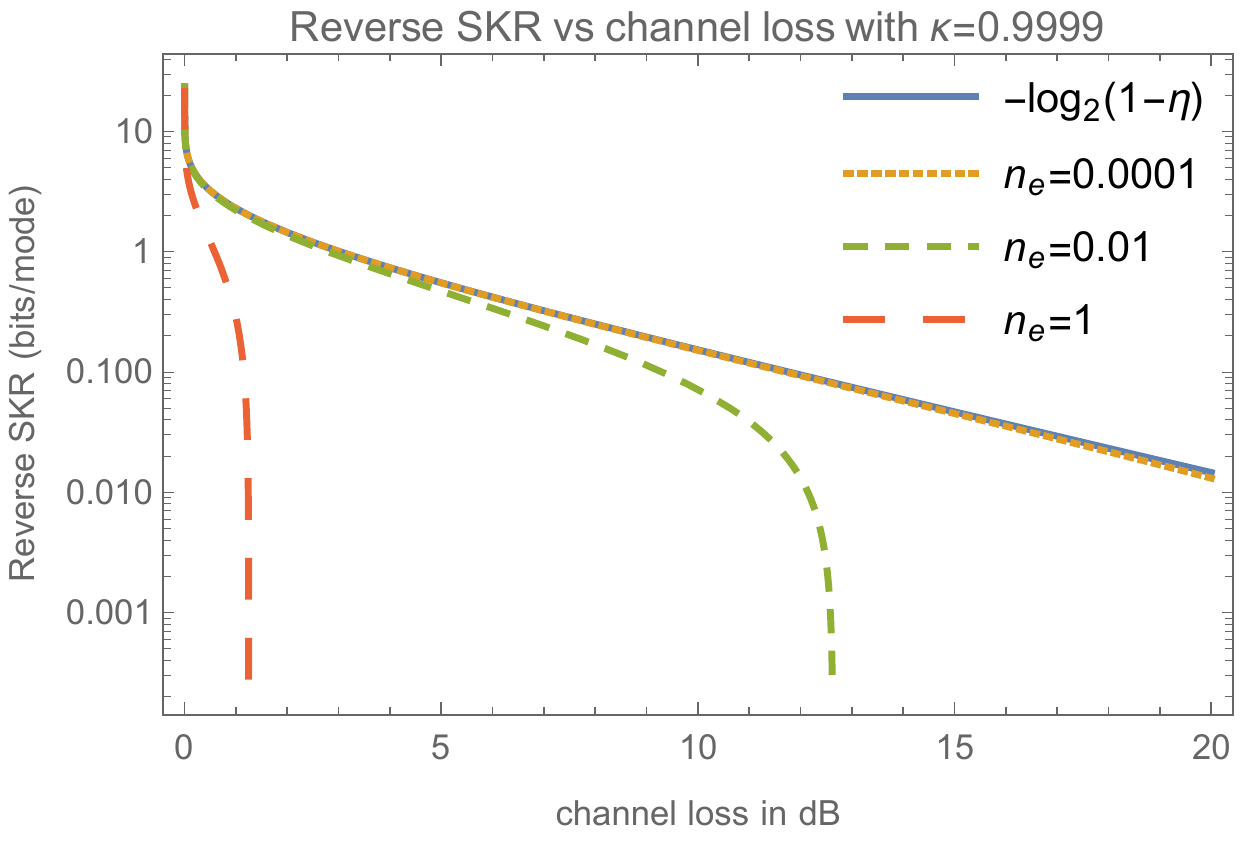}\\
\includegraphics[height=0.8pc]{a.pdf}}\\
{\includegraphics[height=11.8pc]{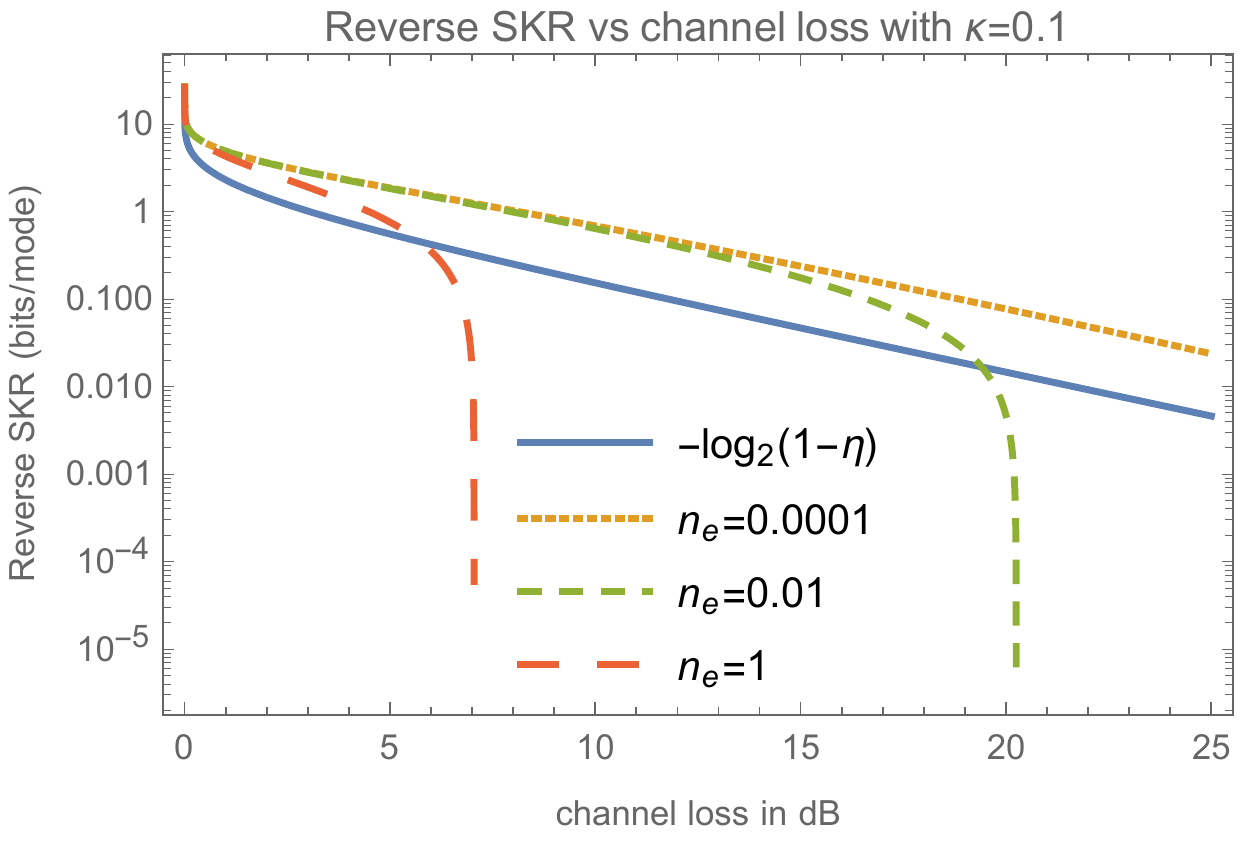}\\
\includegraphics[height=0.8pc]{b.pdf}}
\caption{Achievable rate as a function of  channel loss in dB with different choices of $\kappa$. \label{SKRvLossK1}}
\end{figure}

\begin{figure}[H] 
\centering
{\includegraphics[height=11.8pc]{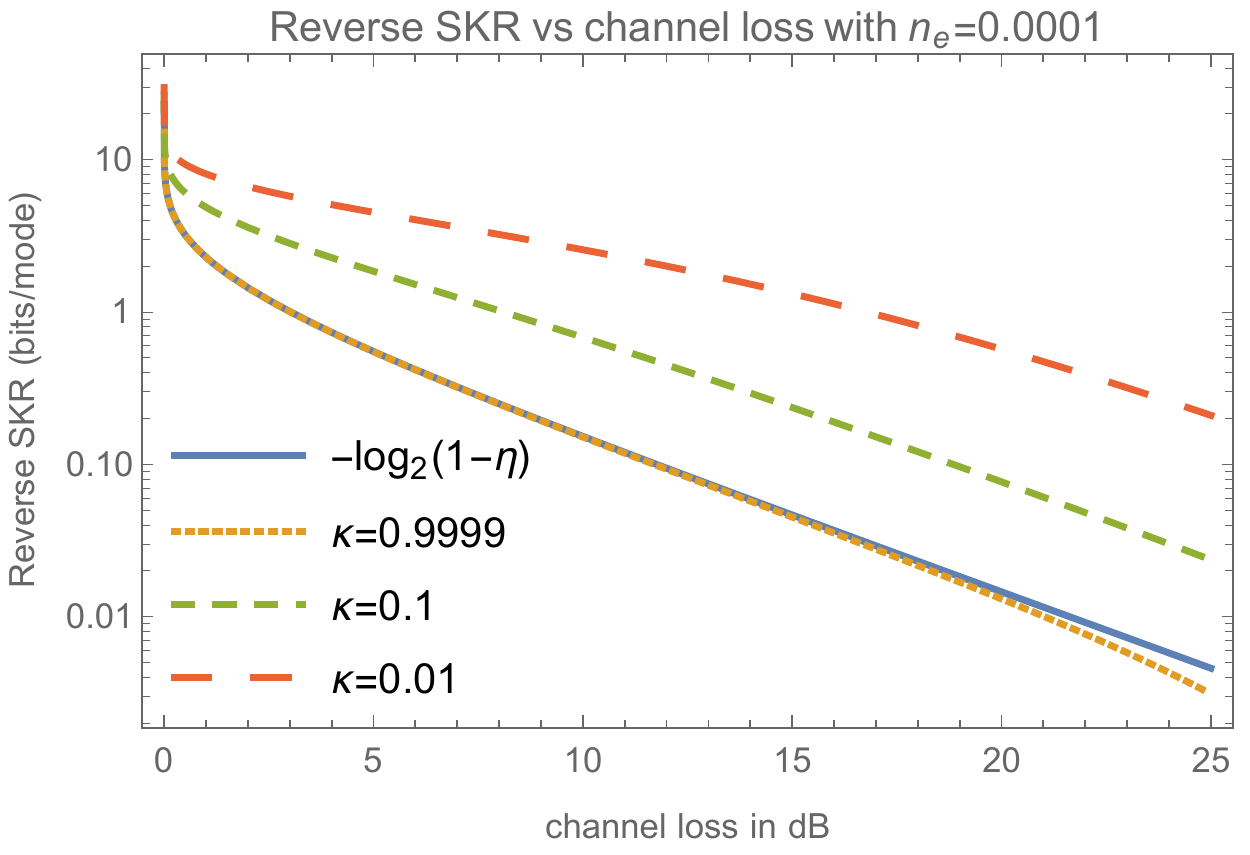}\\
\includegraphics[height=0.8pc]{a.pdf}}\\
{\includegraphics[height=11.8pc]{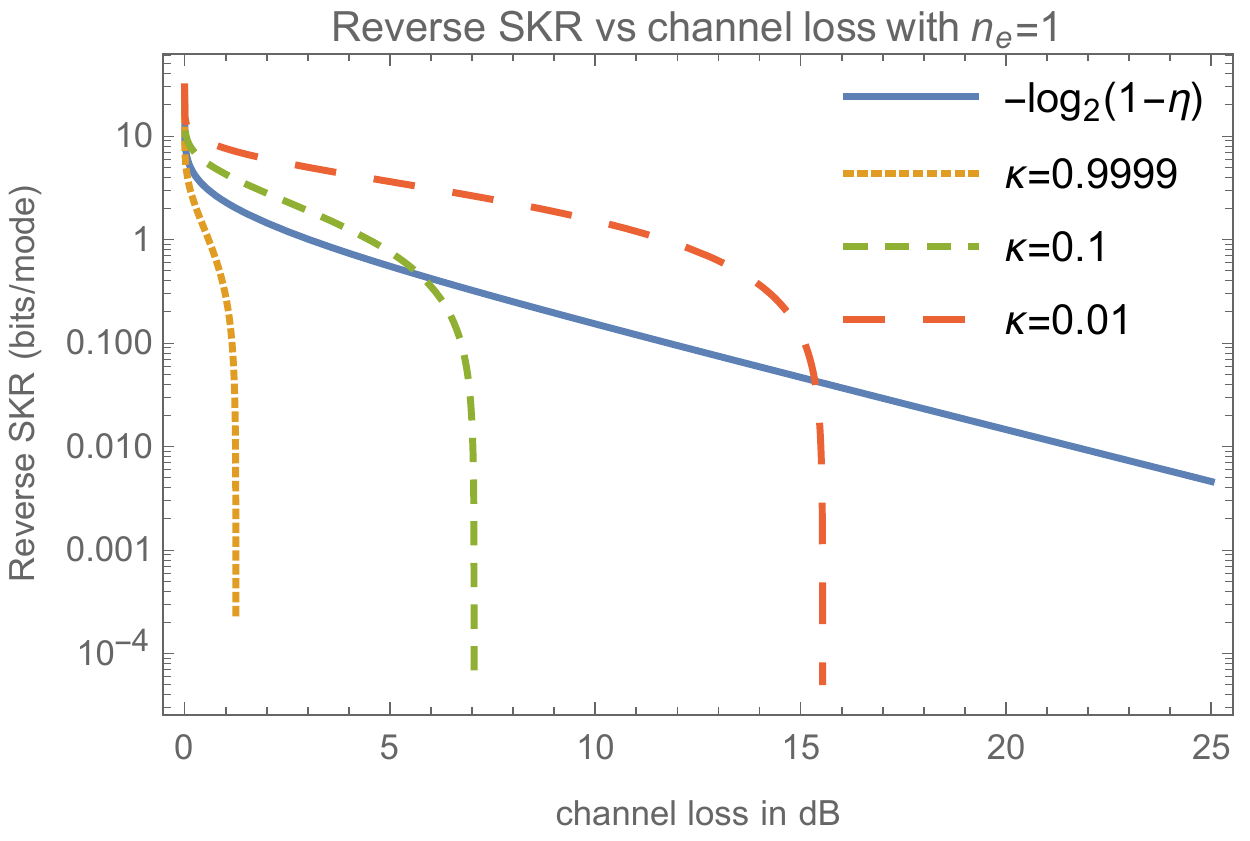}\\
\includegraphics[height=0.8pc]{b.pdf}}
\caption{Achievable rate as a function of  channel loss in dB with different values of $n_e$. \label{SKRvLossn0}}
\end{figure}

Since reverse reconciliation generally gives us a greater transmission range than direct reconciliation (see Fig.~\ref{SKRveta} as transmissivity $\eta$ usually decreases with increasing transmission distance for a given channel), below we show  achievable rates with reverse reconciliation for different parameters. We also plot the capacity of the pure loss channel against the unrestricted Eve, $-\log_2(1-\eta)$, for comparison. In Fig.~\ref{SKRvLossK1}, we show that even though generally with increasing noise the SKR goes down and tends to have a shorter transmission distance, with small $\kappa$ it is still possible to exceed the pure loss unrestricted Eve's  capacity for some values of channel loss. Then we fix the noise $n_e$ and change $\kappa$ in Fig.~\ref{SKRvLossn0}. It is shown that for different values of noise, even when noise is very high, a small $\kappa$ can  improve the achievable rate dramatically.

\subsection{Upper Bounds}\label{UpBR}

In this Section we apply the method discussed in Section \ref{UBs} and Appendix A to get upper bounds for secret key distillation under restricted eavesdropping.

\subsubsection{Pure Loss Channel} First we look at the pure loss case. An upper bound on the secret key distillation capacity under the restricted eavesdropping model considered here follows from the broadcast channel result~\cite[Eq.~(26)]{takeoka2016unconstrained}~\cite[Eq.~(8)]{takeoka2017unconstrained} as:
\begin{equation}
    E_\text{R}(B;AF)_\phi=\log_2\left(\frac{1-\eta_F}{1-\eta_B-\eta_F}\right).
\end{equation}

\begin{figure}[htbp]
	\centering{\includegraphics[width=0.51\textwidth ]{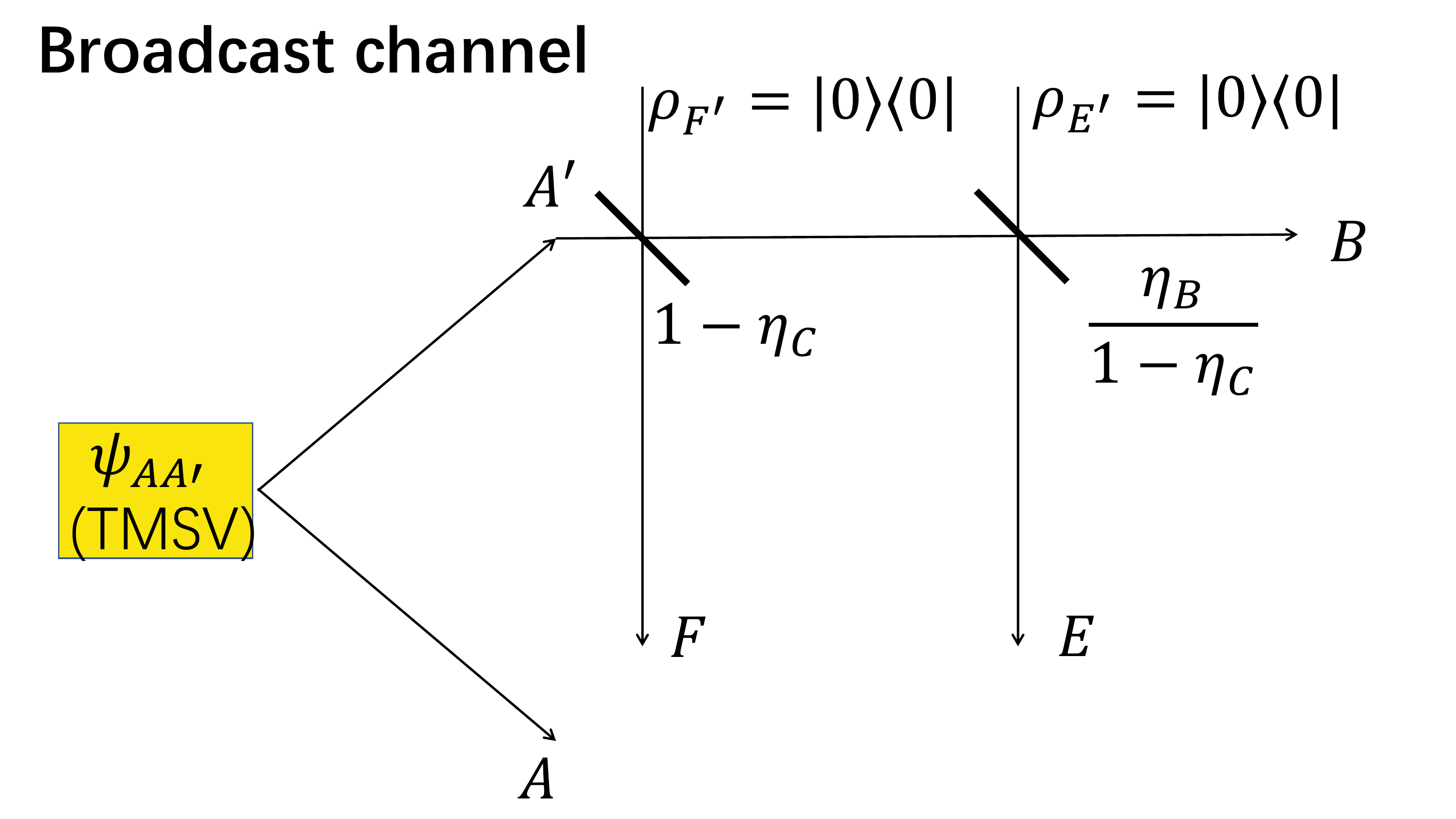}}
	\caption{Broadcast channel shown in \cite{takeoka2016unconstrained,takeoka2017unconstrained}, where a single sender sends information to receivers $F$, $E$ and $B$ through different lossy channels. Here the signal state is sent out from $A'$ while only vacuum states are injected from $F'$ and $E'$. This can be viewed as equivalent to our model if we consider mode $F$ as an inaccessible system, and $E$ as the eavesdropper Eve. \label{broadcast}} 
\end{figure}

Here the notation $B;AF$ means that the closest separable state for the relative entropy entanglement calculation is a state separating system $B$ from systems $AF$. In Ref.~\cite{takeoka2016unconstrained,takeoka2017unconstrained}, the key to obtaining the above bound was the different physical realizations of the same broadcast channel, one of them being as shown in Fig.~\ref{broadcast}. Since only vacuum states are injected from $F'$ and $E'$ in Fig.~\ref{broadcast}, it is equivalent to our model in Fig.~\ref{fig1} with 
$\eta_B=\eta$ and $\eta_C=(1-\kappa)(1-\eta)$. Thus, the upper bound expression for our restricted eavesdropping case is obtained as
\begin{equation}
    \max\{K_\rightarrow,K_\leftarrow\}\le E_\text{R}(B;AF)=\log_2\left(\frac{\eta+\kappa(1-\eta)}{\kappa(1-\eta)}\right).\label{UpBREE}
\end{equation}

\begin{figure}[H] 
\centering{\includegraphics[height=18pc]{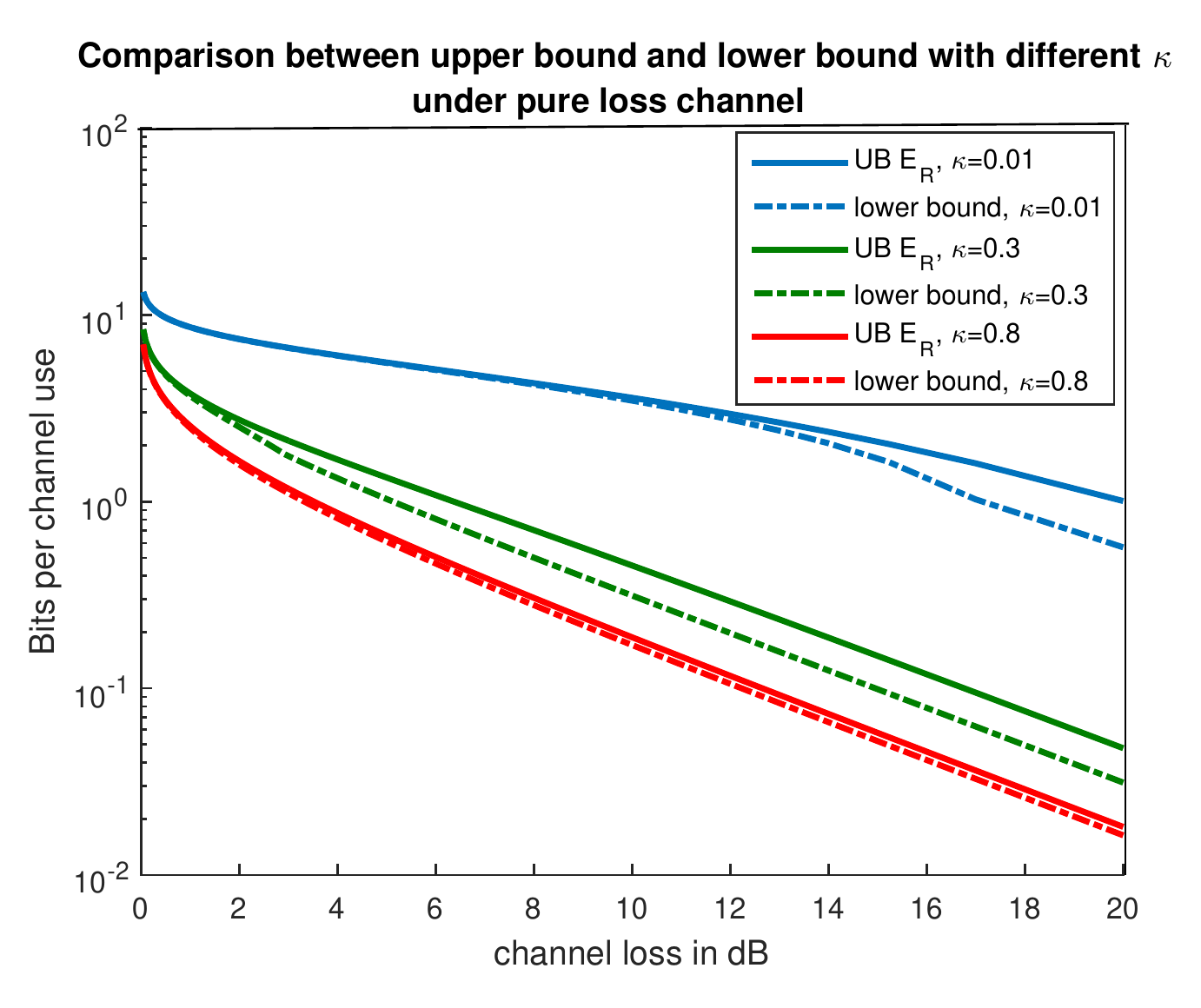}}
\caption{Relative entropy of entanglement upper bounds (UB $E_{R}$) and lower bounds with different values of thermal noise. 
\label{REEdBPL}}
\end{figure}

In Fig.~\ref{REEdBPL} we apply Eq.~(\ref{UpBREE}) to compare the upper bound with the lower bound for different values of $\kappa$. 
We plot the relative entropy entanglement upper bound and the lower bound for three $\kappa$ values, denoted by different colors. Here the lower bound is taken as the maximum of direct  and reverse reconciliation rates.

Here we can see that when $\kappa$ is close to 1, the upper bound almost matches the lower bound. And they match each other when $\kappa=1$ which corresponds to the unrestricted case~\cite{pirandola2017fundamental}. However when $\kappa$ decreases the upper bound becomes looser, but still highly constrain the region in which the capacity for secret key distillation can lie.

One interesting thing to see from Fig.~\ref{REEdBPL} is that the region where direct reconciliation gives a higher rate than reverse reconciliation has a large overlap with the region where the upper bound and lower bound are closest to each other. For example, when $\kappa=0.01$ in Fig.~\ref{REEdBPL} the upper bound and lower bound diverge from each other close to the point where direct reconciliation starts to give a lower rate than reverse reconciliation. Another observation from the plot is that when $\kappa$ decreases the lower bound tends to decrease slower with increasing channel loss at least when channel loss is low.

\subsubsection{Thermal Noise Channel} 
\begin{figure}[H] 
\centering{\includegraphics[height=18.5pc]{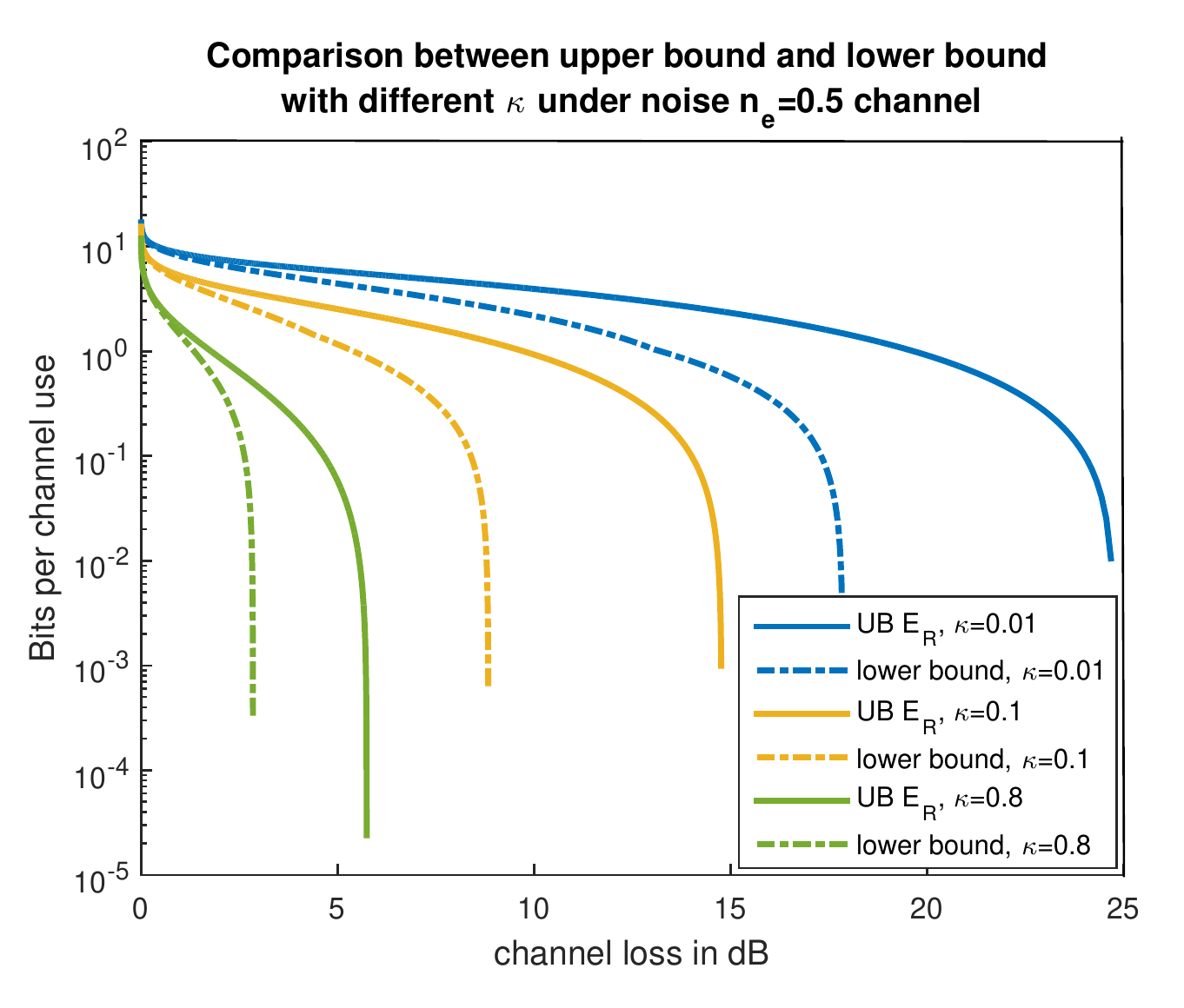}}
\caption{Relative entropy of entanglement upper bounds (UB $E_{R}$) and lower bounds with different values of $\kappa$. 
\label{REEdBne}}
\end{figure}

Next, we look at the thermal noise channel. We plot the relative entropy of entanglement upper bound $E_\textrm{R}(B;AF)$ calculated as described in Appendix~\ref{appdREE} along with the lower bounds of Eqs.~(\ref{DireLBHL}) and (\ref{ReveLBHL2}) in Fig.~\ref{REEdBne}. 
Here we can see that when noise is introduced into the channel the upper bound becomes looser compared to the pure loss channel, which is similar to what was found for unrestricted eavesdropping~\cite{pirandola2017fundamental}. 
These gaps between our upper bounds and lower bounds narrow the search region for this problem's capacity.

\section{Comparison between CV and DVQKD under Restricted Eavesdropping}\label{comp}

In this Section, we present a comparison between achievable rates against a restricted Eve for the Gaussian-modulated CVQKD protocol (with coherent states, heterodyne detection and reverse reconciliation) and the corresponding lower bounds for the DV protocol DS-BB84 ($K_\textrm{DS-BB84}$) that are derived in Appendix~\ref{APB}. 
For better illustration, we also include the upper bounds from Sec.~\ref{UpBR}.

\begin{figure}[H] 
\centering{\includegraphics[height=14pc]{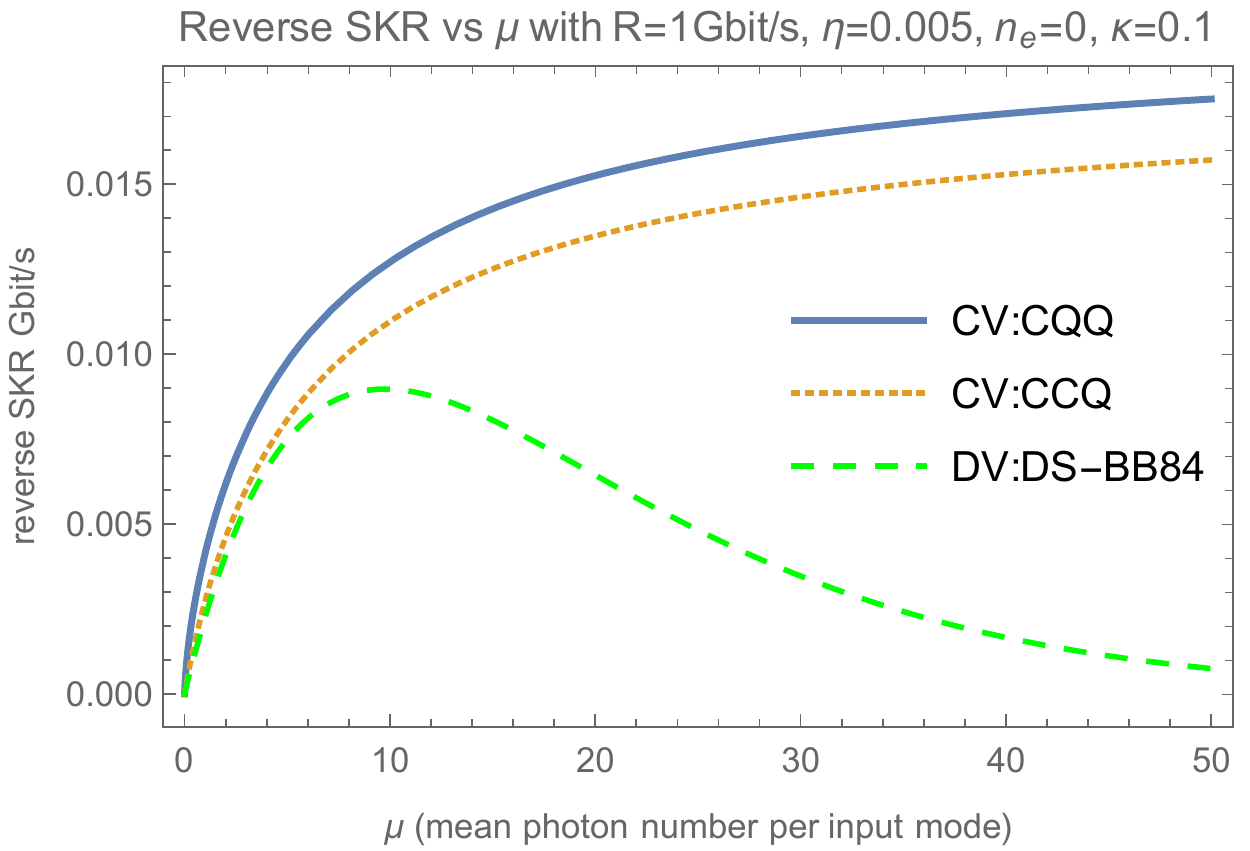}\\
\includegraphics[height=0.8pc]{a.pdf}}\\
{\includegraphics[height=14pc]{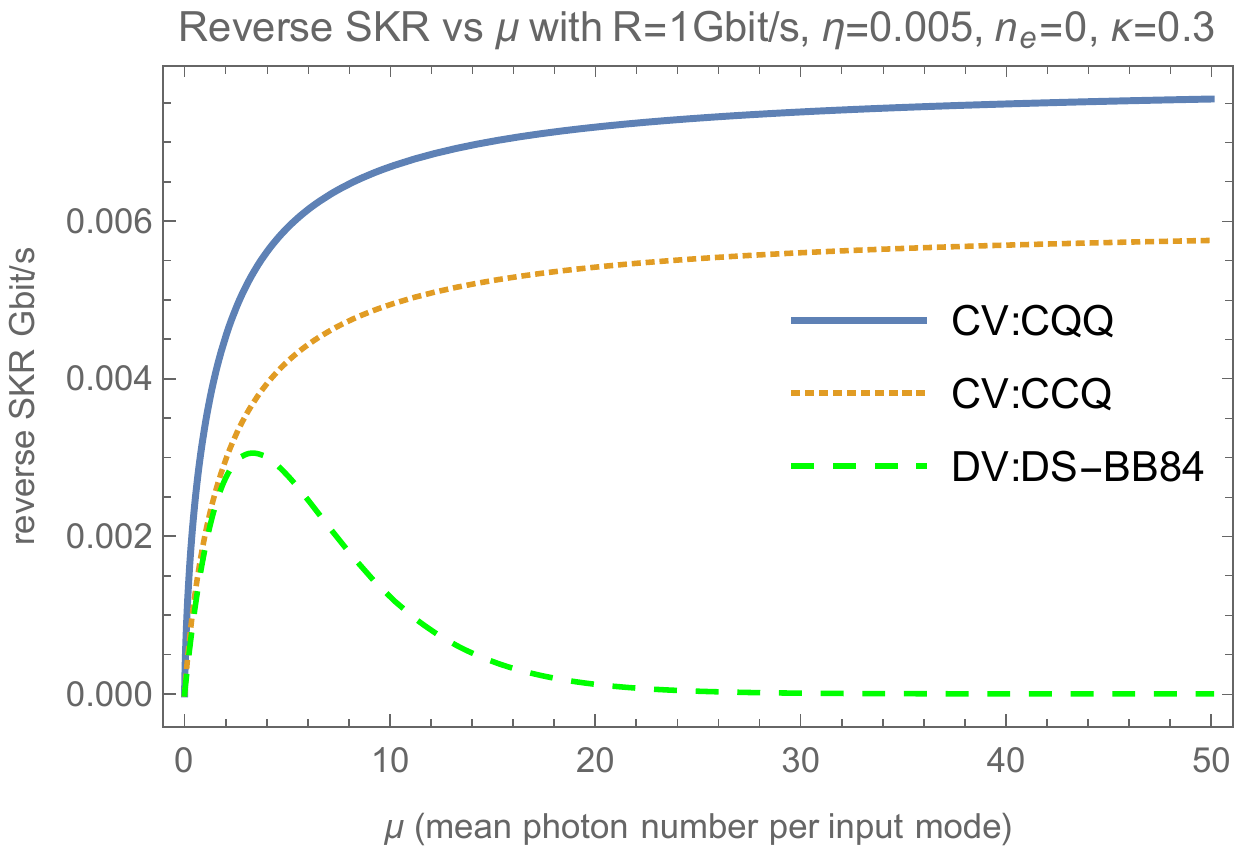}\\
\includegraphics[height=0.8pc]{b.pdf}}
\caption{Comparison of achievable rates for secret key distillation from TMSV state with heterodyne detection and reverse reconciliation vs. DS-BB84 protocol over a pure loss channel. We choose different values of $\kappa$ to show the difference between two protocols. For DS-BB84 (see Appendix~\ref{APB}) we set $\eta_E=\kappa(1-\eta)$ and $f_L=1$.\label{ComPL1}}
\end{figure}

First we look at the pure loss channel. In Fig.~\ref{ComPL1}, we plot the CV achievable rate with reverse reconciliation ($K_\textrm{CQQ}$) and the CV reverse secure key rate when heterodyne detection is performed on both communication sides ($K_\textrm{CCQ}$), and compare with $K_{\textrm{DS-BB84}}$. Here the CCQ case corresponds to actual CVQKD, where Alice sends Gaussian modulated coherent states and Bob performs heterodyne detection. For generality, in a thermal noise channel (when $n_e=0$ this goes back to the pure loss channel):
\begin{align}
K_{\textrm{CCQ}} &\geq I(X;Y)_\omega-I(Y;ER)_\omega\label{CCQline1}\\
&=H(\rho^X)-H(\rho^{ER})-\int_y\!dy\,p(y)(H(\rho_y^X)-H(\rho_y^{ER})),
\end{align}
the state $\omega^{XYER}$ is
\begin{equation}
\omega^{XYER}=\int_x\int_y\!dx\,dy\,p(x)p(y)|x\rangle\langle x|^X\otimes|y\rangle\langle y|^Y\otimes\rho^{ER}_{xy},
\end{equation}

Here $H(\rho^X)$ and $H(\rho_y^x)=H(p(y|x))$ are classical differential entropy of corresponding probability distribution and conditional probability distribution. In Fig.~\ref{ComPL1}, we can see in the pure loss channel for any value of input power, the CV reverse reconciliation scheme generally has rates that are higher than  DS-BB84. Also in the analysis of DS-BB84 there is an optimal input photon number which is why we see the peak in those green curves whereas the optimal input photon number in the CV scheme is infinity, and hence the rate keeps increasing with increasing input photon number.

\begin{figure}[H] 
\centering{\includegraphics[height=13pc]{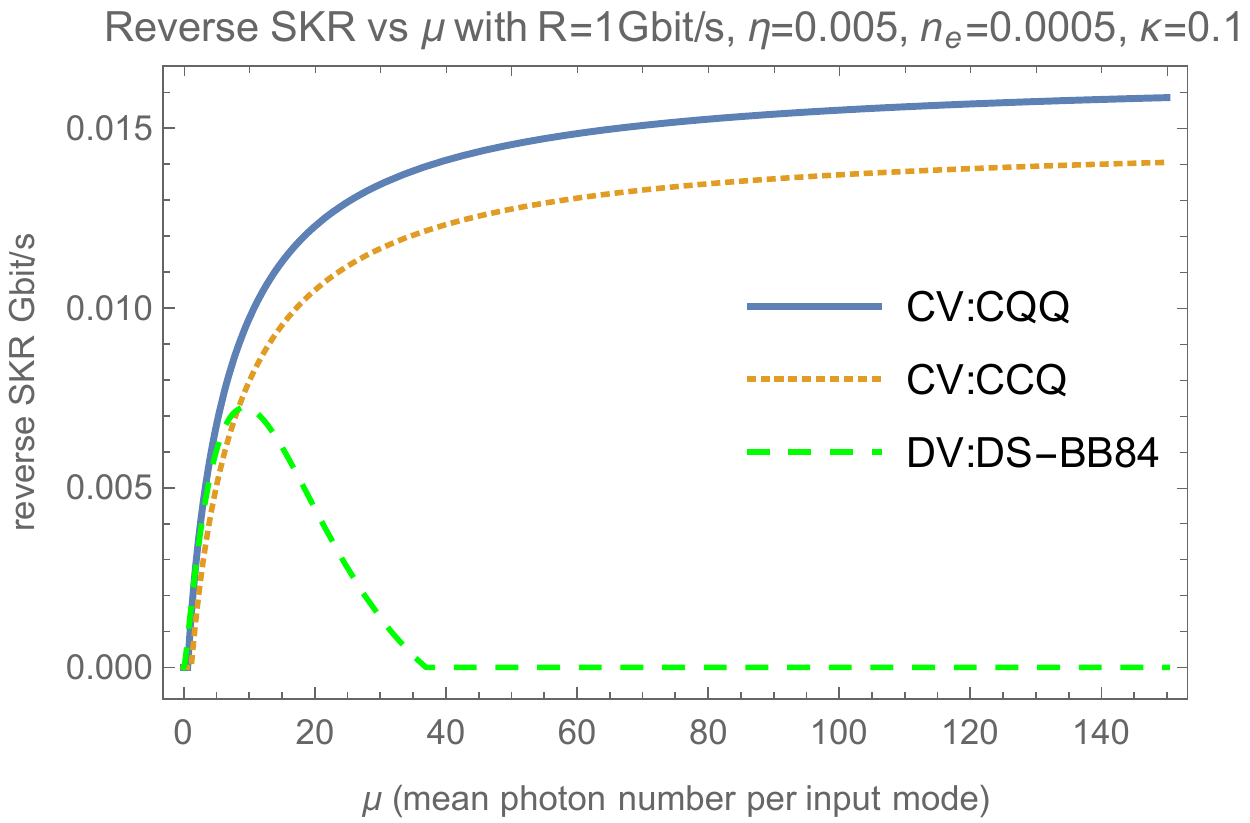}\\
\includegraphics[height=0.8pc]{a.pdf}}\\
{\includegraphics[height=13pc]{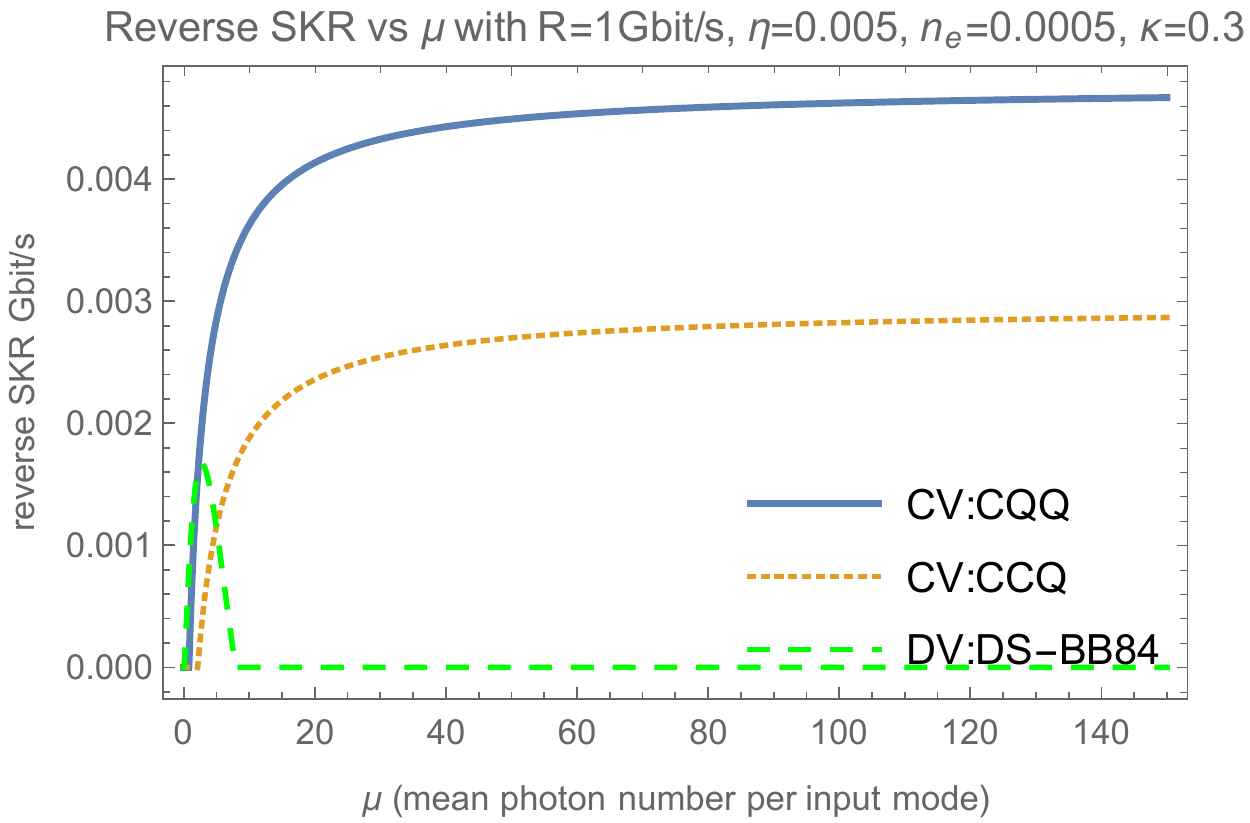}\\
\includegraphics[height=0.8pc]{b.pdf}}
\caption{Comparison between SKRs from: 1. Entanglement-based secret key distillation scheme with heterodyne detection and reverse reconciliation (CQQ), 2. Gaussian modulation reverse reconciliation CVQKD with heterodyne detection (CCQ), 
2. DS-BB84 protocol with similar restricted eavesdropping case.  Here noise $n_e=0.0005$. For DS-BB84 (see Appendix~\ref{APB}) we set $n_d=n_e$, $\eta_E=\kappa(1-\eta)$ and $f_L=1$.\label{Comp2k01}}
\end{figure}

Next we look at the thermal noise channel. In Fig.~\ref{Comp2k01}, we can see that both $K_\textrm{CQQ}$ and $K_\textrm{CCQ}$ increase with increasing $\mu$, whereas the DV rate first increases then decreases. This shows us that the optimal input photon number is different between two schemes. We can also see that although in the pure loss channel CVQKD always have higher SKR than DS-BB84, in a thermal noise channel DS-BB84 can have higher rate for certain channel parameters and input mean photon number.

\begin{figure}[H] 
\centering{\includegraphics[height=13pc]{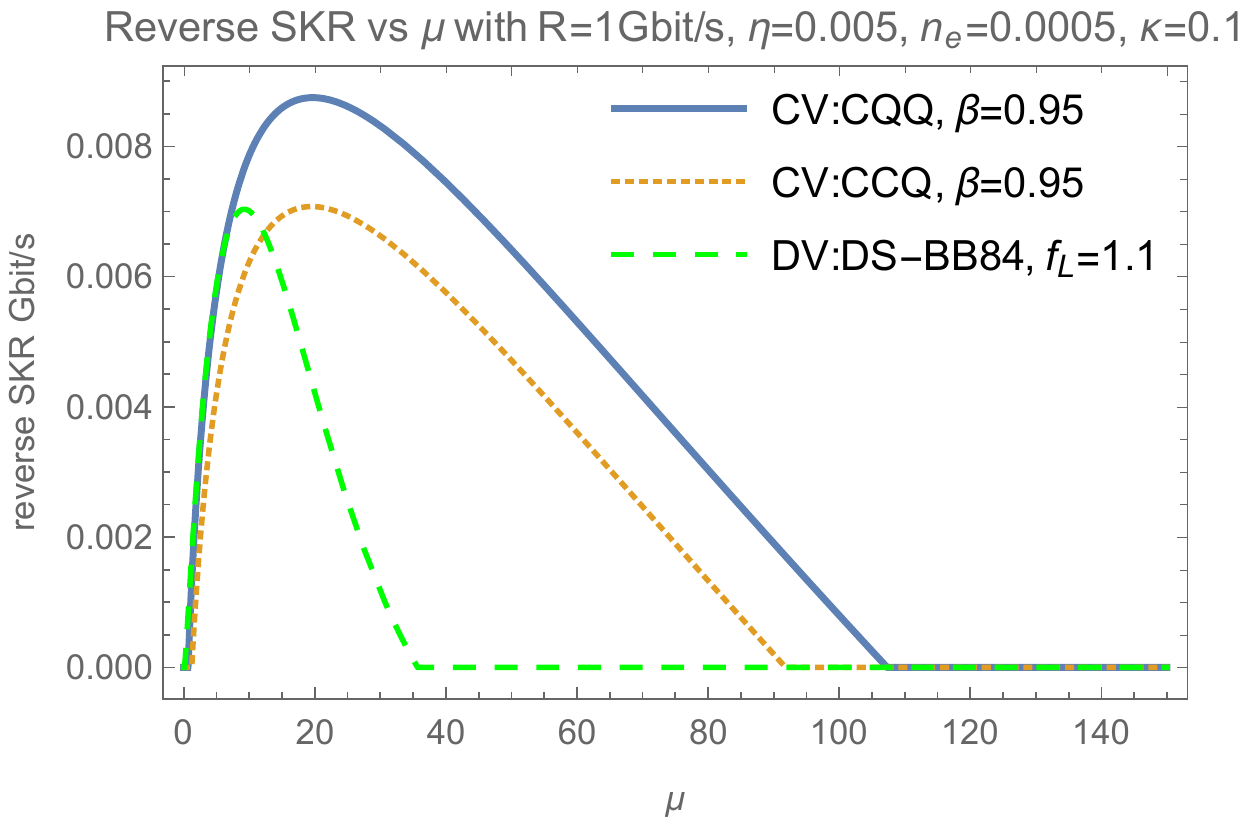}\\
\includegraphics[height=0.8pc]{a.pdf}}\\
{\includegraphics[height=13pc]{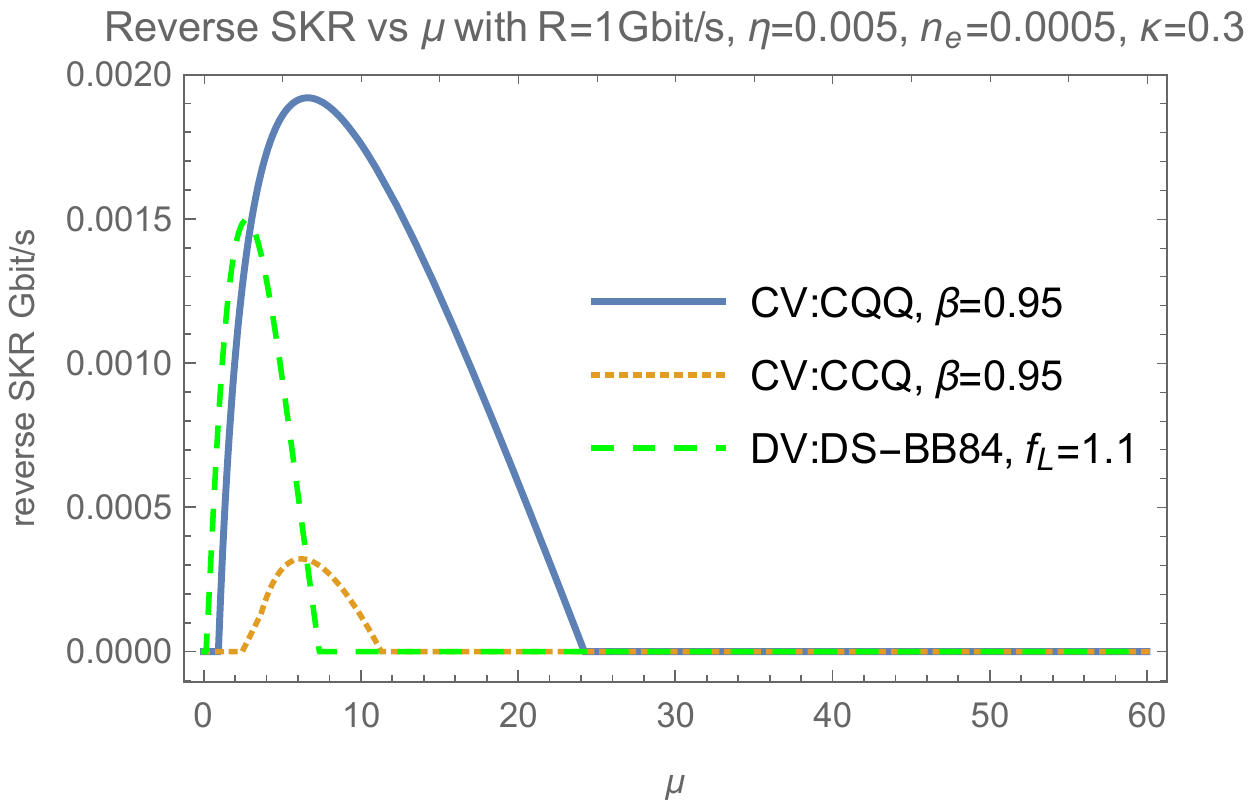}\\
\includegraphics[height=0.8pc]{b.pdf}}
\caption{Comparison between SKRs from: 1. Entanglement-based secret key distillation scheme with heterodyne detection and reverse reconciliation (CQQ) with reconciliation efficiency $\beta=0.95$, 2. Gaussian modulation reverse reconciliation CVQKD with heterodyne detection (CCQ) with reconciliation efficiency $\beta=0.95$, 
2. DS-BB84 protocol with similar restricted eavesdropping case with $f_L=1.1$. (See Eq. (\ref{Iab}))  Here noise $n_e=0.0005$. For DS-BB84 (see Appendix~\ref{APB}) we set $n_d=n_e$ and $\eta_E=\kappa(1-\eta)$.\label{Comp3k01}}
\end{figure}

However, In practical implementation it is not realistic to assume that Alice and Bob can extract all $I(X;Y)$ information through information reconciliation stage. So here in Fig.~\ref{Comp3k01} we take the reconciliation efficiency $\beta$ into consideration and replot the curves from Fig.~\ref{Comp2k01}. We can see now for CVQKD we do have a finite optimal input mean photon number, which is beneficial for implementation since infinite mean photon number is impractical. Instead of Eq.~(\ref{CCQline1}) we use:
\begin{align}
K_{\textrm{CCQ}} &\geq \beta I(X;Y)_\omega-I(Y;ER)_\omega\\
&=\beta H(\rho^X)-H(\rho^{ER})\nonumber\\
&-\int_y\!dy\,p(y)(\beta H(\rho_y^X)-H(\rho_y^{ER})),
\end{align}
the state $\omega^{XYER}$ is
\begin{equation}
\omega^{XYER}=\int_x\int_y\!dx\,dy\,p(x)p(y)|x\rangle\langle x|^X\otimes|y\rangle\langle y|^Y\otimes\rho^{ER}_{xy},
\end{equation}

\begin{figure}[H] 
\centering{\includegraphics[height=17pc]{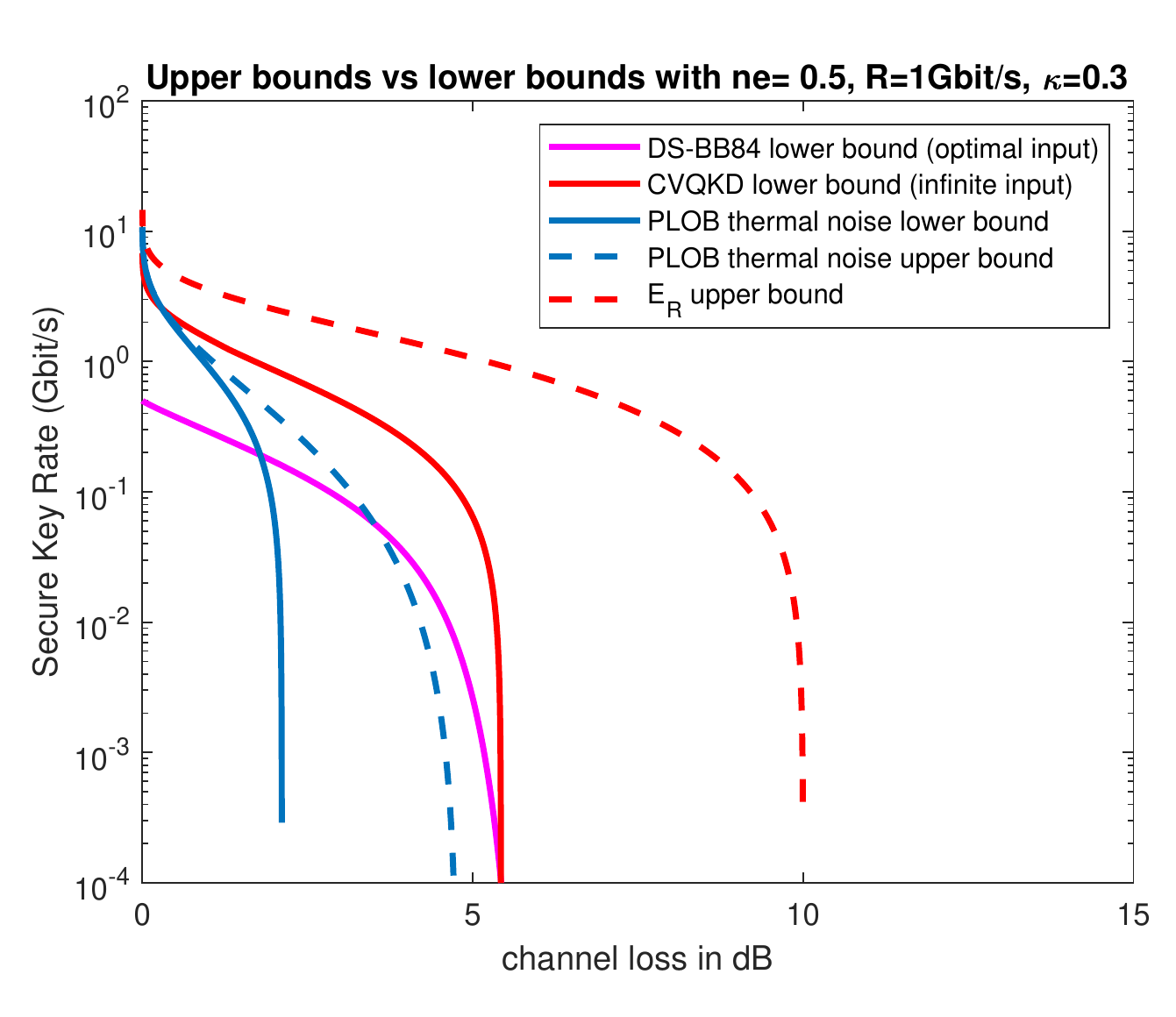}}
\caption{Comparison between CVQKD SKR and optimized DS-BB84 SKR. 
Upper bound and unrestricted Eve's capacity are given as reference lines.\label{Comp2n0}}

\end{figure}

We also plot the comparison between CVQKD and DS-BB84 in Fig.~\ref{Comp2n0}, where the rate of DS-BB84 is optimized over input photon number. 
The rate for CVQKD is for optimal, i.e., infinite, input photon number (assuming $\beta=1$ and $f_L=1$). We also plot the relative entropy of entanglement upper bound calculated in this paper together with the PLOB lower bound and upper bound for thermal noise channel~\cite{pirandola2017fundamental} for reference. It shows the comparison between optimized DV and CV rates with thermal noise $n_e=0.5$ as a function of the channel loss ($n_d=n_e$). We can see that although generally CVQKD has a higher rate, it appears that its achievable rate doesn't necessarily offer a longer distance. Also it can be shown that by assuming such restriction we are getting a higher capacity region. 

\section{Conclusions and Summary}\label{concl}

In summary, we showed lower bounds (achievable rates) for secret key distillation under restricted eavesdropping across pure loss and thermal noise channels based on heterodyne detection. We showed that putting a reasonable restriction on Eve can increase the key rate and extend the transmission range under the same channel conditions. We also showed that unlike the case in unrestricted eavesdropping model, in restricted eavesdropping model direct reconciliation has a potential of providing higher secure key rate under certain channel parameters. Furthermore, we calculated upper bounds using the relative entropy of entanglement for both pure loss and thermal noise channel, and showed that the upper bound is fairly close to the achievable rates with heterodyne detection, for pure loss channel even nearly being the capacity under the restricted eavesdropping model. We also showed a comparison of achievable rates between Gaussian modulation CVQKD and DS-BB84 protocol under restricted eavesdropping with perfect or imperfect reconciliation. All our results capture how the key rates and the transmission distances can increase with the assumption of restricted eavesdropping.

One possible avenue for future work is to find the better detection scheme than heterodyne detection or tighter upper bounds under restricted eavesdropping.

\section{Acknowledgements}
KPS thanks Mark M. Wilde for valuable discussions. The authors thank Tim Ralph for pointing us to a related work~\cite{hosseinidehaj2018optimal}. The authors also gratefully acknowledge financial support from General Dynamics on a QIS IRAD project, and the Office of Naval Research for funding under the Communications and Networking with Quantum Operationally-Secure Technology for Maritime Deployment (CONQUEST) program under contract number N00014-16-C2069, and a MURI program on free-space QKD under contract number N00014-13-1-0627.

\bibliographystyle{IEEEtran}
\bibliography{paper1}

\begin{thebibliography}{10}
\bibitem{AppBfootnote}
In practice, the quantum bit error rate is a measured quantity. So, in order to
  quantify the DS-BB84 protocol's SKR, we will evaluate $\Pr(B_e)$ on the
  presumption that Eve does not affect Alice's single-photon transmissions, in
  which case $\Pr(B_e)$ is due entirely to the presence of dark counts.
\end{thebibliography}

\appendices

\section{Calculating the Relative Entropy of Entanglement Upper Bound}\label{appdREE}
The covariance matrix $V$ of an $N-$mode Gaussian state $\rho$ is the real symmetric matrix defined by
\begin{equation}
V_{ij}=\operatorname{Tr}\left(\rho\{(\hat{r}_i-d_i),(\hat{r}_j-d_j)\}\right),
\end{equation}
where "\{\}" denotes the anticommutator $\{\hat{A},\hat{B}\}=\hat{A}\hat{B}+\hat{B}\hat{A}$. $\hat{r}$ represents the grouped quadrature operators of the $N$ modes involved
\begin{align}
\hat{r}&=\left(\hat{r}_1,...,\hat{r}_{2N}\right)^T\\
&=(\hat{x}_1,...,\hat{x}_N,\hat{p}_1...,\hat{p}_N)^T,
\end{align}
and $d$ denotes the quadrature means
\begin{equation}
d=\operatorname{Tr}\left(\rho\hat{r}\right).
\end{equation}
A valid quantum covariance matrix satisfies the Heisenberg uncertainty principle given by
\begin{align}
V-i\Omega\geq 0,\label{HUP}
\end{align}
where
\[
\Omega=\begin{pmatrix}0 & 1\\
-1 & 0
\end{pmatrix}\otimes I_{N\times N}
\]

Calculating the $E_{\textrm{R}}$ upper bounds for the pure loss and thermal noise channels involves finding the closest separable state to the Gaussian entangled state shared across the channel when one share of a TMSV input is transmitted through the channel. This is accomplished using the positive partial transpose (PPT) criterion, which is a necessary and sufficient condition for separability of 1-vs-$n-$mode bipartite Gaussian states~\cite{lami2018gaussian, Simon00}. The PPT criterion, for a Gaussian state, translates into a condition on its covariance matrix~\cite{lami2018gaussian, Simon00}. For a two-mode covariance matrix $V_{2}$, the condition reads
\begin{align}
V_{2}-i\Omega^{PT}_2\geq0,
\label{PPT2mode}
\end{align}
where 
$\Omega^{PT}_2=\begin{pmatrix}0 & 1\\
-1 & 0
\end{pmatrix}\otimes\begin{pmatrix}1 & 0\\
0 & -1
\end{pmatrix}$.
Equation~(\ref{PPT2mode}) is to be understood as the Heisenberg uncertainty principle after one of the two modes has been time reversed, i.e., $\hat{p}\rightarrow -\hat{p}$ (in this case the second mode), which corresponds to partial transposition.

For a general entangled two-mode covariance matrix of the form 
\begin{equation}
    V_{AB}=\begin{bmatrix}
    a & c_1  \\
    c_1 & b 
\end{bmatrix}
\oplus
\begin{bmatrix}
    a & -c_2  \\
    -c_2 & b
\end{bmatrix},
\label{2ment}
\end{equation}
the closest separable state can be obtained by writing down a covariance matrix
\begin{equation}
    V_2^{\textrm{sep}}=\begin{bmatrix}
    a & c  \\
    c & b 
\end{bmatrix}
\oplus
\begin{bmatrix}
    a & -c  \\
    -c & b
\end{bmatrix},
\label{2msep}
\end{equation}
(where $c$ is an unknown parameter) and determining $c$ using Eq.~(\ref{PPT2mode}) for $V_2^\textrm{sep}$. The value of $c$ such that the smallest eigenvalue of $V_2^{\textrm{sep}}-i\Omega^{PT}_2$ is zero is found to be~\cite{adesso2005gaussian,pirandola2017fundamental}
\begin{equation}
    c=\sqrt{(a-1)(b-1)},\label{copt2}
\end{equation}
and the covariance matrix $V_2^\textrm{sep}$ for the value of $c$ in Eq.~(\ref{copt2}) corresponds to a separable quantum state. The relative entropy between the states corresponding to $V_{AB}$ in Eq.~(\ref{2ment}) and $V_2^\textrm{sep}$ in Eq.~(\ref{2msep}) for $c$ given by Eq.~(\ref{copt2}), when optimized over the input mean photon number gives the upper bounds mentioned in Sec.~\ref{UBs} against traditional unrestricted eavesdropping~\cite{pirandola2017fundamental}.

While the closest separable state gives the tightest relative entropy of entanglement upper bound, a close enough separable state still gives a valid upper bound. We give an $E_{\textrm{R}}$ upper bound for the restricted eavesdropping model by extending the above method. For the three-mode entangled covariance matrix of the form
\begin{align}
V_{ABF}&=\left(
\begin{array}{ccc}
a & c_1 & e \\
c_1 & b & f \\
e & f & d  \\
\end{array}
\right)
\oplus
\left(
\begin{array}{ccc}
a & -c_2 & -e \\
-c_2 & b & f \\
-e & f & d \\
\end{array}
\right)\label{VABF}
\end{align}
that we have in the restricted eavesdropping model, we write down a covariance matrix

\begin{align}
V_3^{\textrm{sep}}&=\left(
\begin{array}{ccc}
a & c & e \\
c & b & f \\
e & f & d  \\
\end{array}
\right)
\oplus
\left(
\begin{array}{ccc}
a & -c & -e \\
-c & b & f \\
-e & f & d \\
\end{array}
\right)\label{PPTM3}
\end{align}
where $c$ is an unknown parameter and numerically solve for $c$ such that the smallest eigenvalue of $V_3^{\textrm{sep}}-i\Omega^{PT}_3$ is zero. Here, 
\begin{equation}
\Omega_3^{PT}=\begin{pmatrix}0 & 1\\
-1 & 0
\end{pmatrix}\otimes\begin{pmatrix}1 & 0 & 0\\
0 & -1 & 0\\
0 & 0 & 1
\end{pmatrix}
\end{equation}
The relative entropy between the states corresponding to $V_{ABF}$ and $V^\textrm{sep}_3$ of Eqs.~(\ref{VABF}) and (\ref{PPTM3}) optimized over the input mean photon number is the upper bound we plot in Secs.~\ref{UpBR} and \ref{comp}. The reason for choosing $V_3^{\textrm{sep}}$ in this form is to simplify the optimization complexity (with only one variable $c$ to be solved) and also because we are only concerned with the separability between modes $A$ and $B$ in the task of distilling secret key (covariance terms other than the covariance between mode $A$ and $B$ are thus left unchanged). In above calculation we are actually studying the separability of mode $AF$ and $B$, which is similar to what was discussed in~\cite{takeoka2017unconstrained} where a second receiver party (Charlie) in broadcast channel assists the communication between Alice and Bob. Also the reason for not considering separating mode $A$ from mode $BF$ is because that would just be studying a quantum channel with different transmissivity with no restrictions on Eve.

\section{Restricted Eve Secure Key Rate for DS-BB84}\label{APB}

In this appendix we derive the SKR for DS-BB84 with a restricted Eve that we used in Sec.~V.  As a prelude, however, we reprise a result from~\cite{SPCDLP09}---using notation that will be convenient for what follows---for an unrestricted Eve.  In both cases, the SKRs are asymptotic-regime results for photon-number splitting attacks.  Moreover, for both cases we assume that Alice and Bob use polarization encoding in which all four polarization states are equally likely.  In particular, Alice uses a weak coherent-state source that transmits signal-state pulses---containing $\mu$ photons per pulse on average at a rate $R$ states/s---over an Alice-to-Bob channel with transmissivity $\eta_c$. She also transmits sufficient decoy states to accurately estimate---in the asymptotic regime---the fraction of her signal pulses that contain single photons.  Bob uses 50--50 active basis selection and a pair of single-photon detectors each with quantum efficiency $\eta_q$ and $n_d$ dark counts on average per pulse interval $1/R$.

When Eve is unrestricted, we must assume that she can interact with \emph{all} the light Alice sends to Bob.  From~\cite{SPCDLP09} we then  have that
\begin{equation}
{\rm SKR} = \max(I_{AB} - I_{AE},0)
\label{DW}
\end{equation}
where $I_{AB}$ is Alice and Bob's Shannon information (in bits/s) and $I_{AE}$ is Alice and Eve's Shannon information (in bits/s).  Alice and Bob's Shannon information obeys
\begin{equation}
I_{AB} = R\Pr(B_1)\{1-f_LH_2[\Pr(B_e)]\}/2
\label{Iab}
\end{equation}
where: $B_1$ is the event that Bob gets a total of 1 click from his two detectors during a signal-pulse interval when Alice and Bob use the same basis, making $\Pr(B_1)$ the probability of a sift event; $B_e$ is the sift event in which Bob's detector click is from the detector associated with the wrong polarization, making $\Pr(B_e)$ the raw-key error probability, a quantity usually called the quantum bit error rate~\cite{AppBfootnote};
\begin{equation}
H_2(p) \equiv -p\log_2(p) -(1-p)\log_2(1-p),
\end{equation}
is the binary entropy function; and $f_L \ge 1$ is Alice and Bob's reconciliation penalty.  In words, this formula states that $I_{AB}$ is the sift rate $R\Pr(B_1)/2$, i.e., the rate at which Alice and Bob choose the same basis \emph{and} Bob gets a total of 1 click from his detectors, multiplied by Alice and Bob's Shannon information per sift event $1-f_LH_2[\Pr(B_e)]$, i.e., the entropy of Alice's transmission to Bob minus the information leaked during reconciliation.
 
To find Alice and Eve's Shannon information rate, consider what happens when Alice transmits an $n$-photon signal pulse.  When $n=0$, Eve gets no information. When $n=1$, Eve gets partial information, but this comes at the expense of her creating errors. When $n\ge 2$, Eve gets complete information by means of a photon-number splitting attack. Consequently,
Alice and Eve's Shannon information rate obeys
\begin{eqnarray}
I_{AE} &=& R\Pr(B_1)\{1-\Pr(A_0\mid B_1)-\Pr(A_1\mid B_1) \nonumber \\
& &\times\,\, (1-H_2[\Pr(B_e \mid A_1B_1)])\}/2,
\label{Iae}
\end{eqnarray}
where $A_n$ is the event that Alice's signal pulse contains $n$ photons.  Here, $I_{AE}$ equals Alice and Bob's sift rate $R\Pr(B_1)/2$ multiplied by the entropy of Alice's transmission to Bob reduced by entropies of the sift events associated with Alice's transmission of a 0 photon signal pulse and those of the sift events associated with Alice's transmission of a 1 photon signal pulse and Eve's interaction therewith.   

The probabilities necessary to instantiate the preceding Shannon information expressions are easily found.  
The probability of a sift event equals the probability of a click on one of Bob's detectors and no click on the other detector:  
\begin{eqnarray}
\Pr(B_1) &=& (1-e^{-(\eta\mu+n_d)})e^{-n_d} \nonumber \\ 
&&+\,\,e^{-(\eta\mu+n_d)}(1-e^{-n_d}),
\label{Pb1}
\end{eqnarray}
where $\eta \equiv \eta_q\eta_c$ is the overall Alice-to-Bob transmissivity (the channel transmissivity times the detector quantum efficiency) and we have used statistically independent Poisson statistics for the counts generated by each detector.  We have that $\Pr(A_0 \mid B_1) = \Pr(A_0B_1)/\Pr(B_1)$, where the joint probability of Alice sending 0 photons and that transmission's resulting in a sift event is 
\begin{equation}
\Pr(A_0B_1) = \Pr(A_0)\Pr(B_1\mid A_0) = 2 e^{-\mu}e^{-n_d}(1-e^{-n_d}).
\end{equation}
Similarly we have that $\Pr(A_1 \mid B_1) = \Pr(A_1B_1)/\Pr(B_1)$, where the joint probability of Alice sending 1 photon and that transmission's resulting in a sift event is 
\begin{eqnarray}
\lefteqn{\Pr(A_1B_1) = \Pr(A_1)\Pr(B_1\mid A_1)}\\[.05in]
&=& \mu e^{-\mu}[\eta e^{-2n_d} + (2-\eta)e^{-n_d}(1-e^{-n_d})].
\end{eqnarray}
Bob's conditional error probability for a sift event, given Alice sent 1 photon, is thus
\begin{eqnarray}
\lefteqn{\Pr(B_e\mid A_1B_1) = \Pr(BeA_1B_1)/\Pr(A_1B_1)} \\[.05in] 
&=& \frac{(1-\eta)e^{-n_d}(1-e^{-n_d})}{\eta e^{-2n_d} + (2-\eta)e^{-n_d}(1-e^{-n_d})},
\end{eqnarray}
where the numerator is the joint probability that Alice has transmitted 1 photon, it results in a sift event, and Bob's single click comes from the detector associated with the wrong polarization.  
Bob's unconditional error probability for a sift event is given by
\begin{eqnarray}
\lefteqn{\Pr(B_e) = } \nonumber \\
&& \hspace{-0.3in}\frac{e^{-(\eta\mu + n_d)}(1-e^{-n_d})}{(1-e^{-(\eta\mu+n_d)})e^{-n_d} + e^{-(\eta\mu+n_d)}(1-e^{-n_d})},
\label{Pbe}
\end{eqnarray}
where the numerator is the probability for a sift event in which Bob's 1 click comes from the detector associated with the wrong polarization.  Substituting these probabilities into Eqs.~(\ref{Iab}), and (\ref{Iae}), and then evaluating Eq.~(\ref{DW}), gives us Alice and Bob's SKR for an unrestricted Eve,
\begin{eqnarray}
\lefteqn{{\rm SKR} = \max\{R\Pr(B_1)\{\Pr(A_0\mid B_1)-f_LH_2[\Pr(B_e)]} \nonumber \\[.05in]
&+&\Pr(A_1\mid B_1)(1-H_2[\Pr(B_e \mid A_1B_1)])\}/2,0\}.
\label{SKRun}
\end{eqnarray}

Alice and Bob's SKR can be significantly better when Eve is restricted, i.e., when she can only interact with a fraction $\eta_E$ of the light transmitted by Alice and that fraction is disjoint from the $\eta_c$ fraction that arrives at Bob's receiver.  It follows that Eve gets no information when Alice's signal pulse contains 0 photons and she gets complete information about every signal pulse from which she collects a photon \emph{and} Bob gets a total of 1 click from his two detectors.  We can now use Eq.~(\ref{DW}) to get Alice and Bob's SKR, with $I_{AB}$ given by Eqs.~(\ref{Iab}), (\ref{Pb1}) and (\ref{Pbe}), and $I_{AE}$ given by
\begin{equation}
I_{AE} = R\Pr(B_1)[1-\Pr(E_0\mid B_1)]/2,
\end{equation}
where $E_0$ is the event that Eve collects none of Alice's photons.  Thus we get 
\begin{equation}
{\rm SKR} = R\Pr(B_1)\{\Pr(E_0\mid B_1)-f_LH_2[\Pr(B_e)],0\}/2,
\label{SKRexcl}
\end{equation}
for the DS-BB84 system with a restricted Eve, where
\begin{equation}
\Pr(E_0\mid B_1)  = e^{-\eta_E\mu},
\end{equation}
because $B_1$ and $E_0$ are statistically independent events.  

\begin{figure}[hbt]
\includegraphics[width=3.5in]{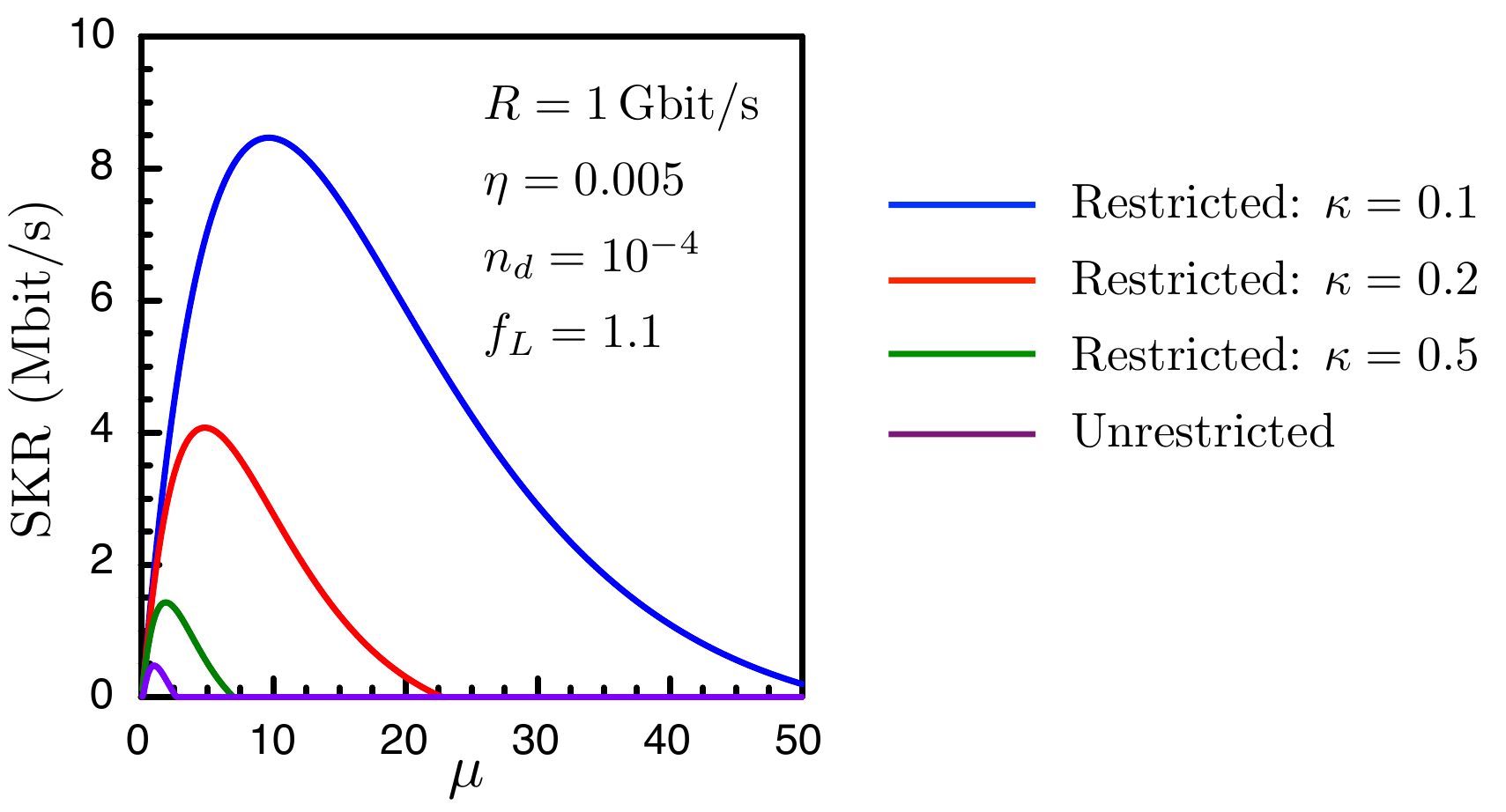}
\caption{SKRs for unrestricted and restricted Eves plotted versus the average photon number, $\mu$, of Alice's signal-state transmissions. \label{SKR_vs_mu}}
\end{figure}

Figure~\ref{SKR_vs_mu}, obtained using the parameters given in Table~\ref{Fig1_params}, illustrates the very significant SKR increase afforded by Eve's being restricted in her access to Alice's transmitted light, i.e., when $\kappa \equiv \eta_E/(1-\eta) \ll 1$.  
\begin{table}[thb]
\begin{tabular}{|c|c|c|}\hline
Parameter & Symbol & Value\\ \hline
Alice's signal-state transmission rate & $R$ & 1\,Gbit/s\\
Overall Alice-to-Bob transmissivity & $\eta$ & 0.005\\ 
Alice's signal-state average photon number & $\mu$ & see Fig.~\ref{SKR_vs_mu}\\ 
Alice-to-Eve transmissivity & $\eta_E=\kappa(1-\eta)$ & see Fig.~\ref{SKR_vs_mu}\\ 
Average detector dark-count per bit interval & $n_d$ & $10^{-4}$\\
Reconciliation penalty & $f_L$ & 1.1\\
\hline
\end{tabular}\\
\caption{Parameter values used for the SKR comparison in Fig.~\ref{SKR_vs_mu}\label{Fig1_params}}.
\end{table}

\end{document}